\documentclass[11pt,a4paper]{article}
\pdfoutput=1
\usepackage{jheppub}
\usepackage{rotating}
\usepackage{array}
\usepackage{amsmath}
\usepackage[normalem]{ulem}
\usepackage{slashed}
\usepackage{booktabs}
\usepackage[pdftex,table,dvipsnames]{xcolor}
\usepackage{units} 
\usepackage{multirow}
\usepackage{url}
\usepackage{parskip}
\usepackage[all]{nowidow}
\usepackage{placeins}

\usepackage{xspace}
\usepackage{subfig}
\usepackage{color}

\usepackage{dsfont}
\usepackage{soul}
\usepackage{hyperref}

\allowdisplaybreaks

\def \ff {f\!f}
\newcommand{\me}{\ensuremath{\slashed{E}}\xspace}

\preprint{NIKHEF-2022-007}

\title{Global Analysis of the ALP Effective Theory}

\author[1,2]{Sebastian Bruggisser,}
\author[1]{Lara Grabitz,}
\author[1,3,4]{and Susanne Westhoff\,}

\affiliation[1]{Institute for Theoretical Physics, Heidelberg University, Heidelberg, Germany}
\affiliation[2]{Department of Physics and Astronomy, Uppsala University, Uppsala, Sweden}
\affiliation[3]{Institute for Mathematics, Astrophysics and Particle Physics, Radboud University,
Nijmegen, The Netherlands}
\affiliation[4]{NIKHEF Theory Group, Amsterdam, The Netherlands\\}

\emailAdd{bruggisser@thphys.uni-heidelberg.de}
\emailAdd{susanne.westhoff@ru.nl}

\abstract{We perform a global fit of the effective Lagrangian for axion-like particles (ALPs) to data. By combining LHC observables from top physics, dijet and di-boson production with electroweak precision observables, we resolve the full parameter space of ALPs with flavor-universal couplings. Using the renormalization group to evolve the effective ALP couplings to low energies allows us to investigate the impact of flavor observables on the global analysis. We show that resonance searches in $B\to K$ meson decays significantly enhance the sensitivity to ALPs with sub-GeV masses. The lifetime of the ALP plays a crucial role in resolving the multi-dimensional parameter space with searches for prompt, displaced and invisible ALP decays. Our analysis points out the differences in probing an effective theory with new light particles, compared to scenarios with only non-resonant effects of heavy particles at low energies, as in the Standard Model Effective Field Theory.}

\begin{document}

\maketitle

\flushbottom

\section{Introduction}
\label{sec:introduction}
Effective field theories are an elegant tool to explore extensions of the Standard Model (SM) that involve different scales. The effective theory of axion-like particles (ALPs)~\cite{Georgi:1986df} describes a class of a theories with two scales: the mass of the ALP and the cutoff scale of the effective theory, which is associated with a broken chiral symmetry. For the axion that addresses the CP problem in QCD, both scales are related through the axion decay constant~\cite{Peccei:1977hh,Peccei:1977ur,Weinberg:1977ma,Wilczek:1977pj}. More generally, axion-like particles are predicted as pseudo Nambu Goldstone bosons in many theories with spontaneously broken symmetries.

The separation between the ALP mass and the cutoff scale can be moderate, as for the pions in chiral perturbation theory, or it can span several orders of magnitude in energy, as for the Peccei-Quinn axion in QCD. In this work, we investigate the ALP effective theory in the regime where the ALP mass lies below the scale of QCD confinement and the cutoff scale is above the energy reach of the LHC. Under this assumption, the ALP is a light messenger of a new sector of heavy particles which generate the effective ALP couplings. Additional effects of these heavy particles could directly affect the observables, but are model-dependent. We do not consider them in this work.

ALPs can leave a variety of traces in experiments running at different energy scales. At high-energy colliders, ALPs can modify QCD and electroweak processes through virtual effects~\cite{Alonso-Alvarez:2018irt,Baldenegro:2018hng,Gavela:2019cmq,Carra:2021ycg,Bonilla:2022pxu} and induce signatures with new resonances or missing energy~\cite{Mimasu:2014nea,Jaeckel:2015jla,Bauer:2017ris,Brivio:2017ije,Ebadi:2019gij,Esser:2023fdo,Rygaard:2023vyo}. At flavor experiments, ALPs with masses below a few GeV can be resonantly produced in meson decays~\cite{Batell:2009jf,Freytsis:2009ct,Izaguirre:2016dfi,Choi:2017gpf,Bjorkeroth:2018dzu,Dobrich:2018jyi,Gavela:2019wzg,MartinCamalich:2020dfe,Carmona:2021seb,Bauer:2021mvw,Ferber:2022rsf}, lepton decays~\cite{Bjorkeroth:2018dzu,Bauer:2019gfk,Calibbi:2020jvd}, or directly in $e^+ e^-$ collisions~\cite{Dolan:2017osp,Acanfora:2023gzr}. Sub-GeV ALPs can be also produced and searched for at fixed-target experiments~\cite{Dobrich:2015jyk,Dobrich:2019dxc}. Even lighter ALPs leave imprints on observables in cosmology and astrophysics~\cite{Cadamuro:2011fd,Millea:2015qra,Depta:2020wmr}.

The phenomenology of ALPs in each sector has been explored in detail, see Ref.~\cite{Agrawal:2021dbo} for a summary. Results have often been reported as bounds on one or two specific ALP couplings obtained at a specific energy scale. Such analyses are difficult to interpret in the ALP effective theory, because the effective couplings are scale-dependent and mix under the renormalization group (RG) evolution~\cite{Choi:2017gpf,MartinCamalich:2020dfe,Chala:2020wvs,Bauer:2020jbp}. The RG evolution induces effective ALP couplings at lower scales, even if these couplings are absent at the cutoff scale. Moreover, UV completions of the ALP effective theory are likely to generate several couplings at once at the cutoff scale. In order to obtain a coherent picture of the ALP parameter space, observables at different scales need to be combined in a consistent framework.

We achieve such a coherent picture by performing a global analysis of the ALP effective theory with flavor-universal fermion couplings. We select a set of collider observables that are particularly sensitive to the various ALP couplings in the effective Lagrangian. In each observable, we include all leading ALP effects and express them in terms of the relevant couplings at the cutoff scale of the effective theory. This procedure allows us to perform a joint fit of observables at different energy scales to data. In this way, we resolve the full parameter space of the ALP effective theory. Our results are presented as bounds on the ALP effective Lagrangian at the cutoff scale, which can be directly used to constrain the structure of possible UV completions.

An alternative approach has recently appeared in Ref.~\cite{Biekotter:2023mpd}, where the authors investigate ALP-induced effects on the Standard Model Effective Field Theory (SMEFT) in a global fit. The analysis constrains UV-sensitive loop-induced contributions of ALPs in observables, which form a sub-set of all possible effects.

This article is organized as follows. In Sec.~\ref{sec:alps}, we review the ALP effective theory and set the framework for our analysis. In Sec.~\ref{sec:lhc}, we analyze effects of virtual ALPs in observables at the LHC and LEP. Dijet production, top-antitop production, and four-top production at the LHC resolve the sub-space of ALP couplings to fermions and gluons, $c_{\ff}$ and $c_{GG}$ - Sec.~\ref{sec:top-dijet}. Electroweak precision observables at LEP and di-boson production at the LHC are particularly sensitive to ALP couplings to electroweak gauge bosons, $c_{WW}$ and $c_{BB}$, but also involve the ALP-fermion coupling $c_{\ff}$ - Sec.~\ref{sec:ew}. In combination, the two sectors constrain the full parameter space of the ALP Lagrangian - Sec.~\ref{sec:comb-lhc}. In Sec.~\ref{sec:flavor}, we confront the LHC-LEP fit with flavor observables. At LHCb, BaBar and Belle\,(II), ALPs can be produced through $B\to K$ decays - Sec.~\ref{sec:alps-b-decays}. Signatures with prompt, displaced and invisible ALP decay products probe complementary parameter regions through the lifetime of the ALP - Sec.~\ref{sec:flavor-signals}. We quantify the impact of flavor observables on the global analysis of the ALP effective theory - Sec.~\ref{sec:flavor-impact} - and conclude in Sec.~\ref{sec:conclusions}.
\section{ALP effective theory}
\label{sec:alps}
Axion-like particles are pseudo-scalars whose interactions with Standard Model fields preserve a global shift symmetry $a \to a + c$, where $a$ is the ALP field and $c$ is a constant. Shift-symmetric ALP couplings that preserve the gauge symmetries of the Standard Model occur at mass dimension 5 or higher. At energies above the weak scale $\mu_w$, the ALP couplings can be described by an effective Lagrangian\footnote{We adopt the notation from Ref.~\cite{Bauer:2020jbp}, but express the fermion fields in terms of their mass eigenstates $f$. In the contraction $F_{\mu\nu}\widetilde{F}^{\mu\nu}$ of a field strength tensor $F_{\mu\nu}$ with its dual, $\widetilde{F}^{\mu\nu} = \frac{1}{2}\epsilon^{\mu\nu\rho\sigma}F_{\rho\sigma}$, the summation over gauge indices is implicit.}\footnote{We do not consider potential non-anomalous ALP couplings to weak gauge bosons, which can arise in models with chiral gauge groups~\cite{Quevillon:2019zrd,Bonnefoy:2020gyh}.}
\begin{align}\label{eq:lagrangian}
    \mathcal{L}_{\rm eff}(\mu > \mu_w) & = \frac{1}{2}\left(\partial_\mu a\right)\left(\partial^\mu a\right) - \frac{m_a^2}{2}\,a^2 + \sum_f c_{\ff}(\mu)\,\frac{\partial^{\mu} a}{2f_a}\,(\bar{f} \gamma_{\mu} \gamma_5 f)\\\nonumber
    & + c_{GG}(\mu)\,\frac{a}{f_a}\,\frac{\alpha_s}{4\pi}\,G_{\mu\nu} \widetilde{G}^{\mu\nu}
     + c_{WW}(\mu)\,\frac{a}{f_a}\,\frac{\alpha_2}{4\pi}\,W_{\mu\nu} \widetilde{W}^{\mu\nu} + c_{BB}(\mu)\,\frac{a}{f_a}\,\frac{\alpha_1}{4\pi}\,B_{\mu\nu} \widetilde{B}^{\mu\nu}\,.
\end{align}
This effective theory is valid up to a cutoff scale $\Lambda = 4\pi f_a$, where new particles are expected to complete the theory. In our  analysis, we set $f_a = 1\,$TeV and define all ALP couplings at the cutoff scale $\Lambda = 4\pi\,$TeV,
\begin{align}
c_{\ff} = c_{\ff}(\Lambda),\quad c_{GG} = c_{GG}(\Lambda),\quad c_{WW} = c_{WW}(\Lambda),\quad c_{BB} = c_{BB}(\Lambda)\,.
\end{align}
All couplings are real parameters. ALP couplings $c_{GG}$, $c_{WW}$, $c_{BB}$ to the SM gauge fields are normalized to the gauge couplings $\alpha_s = g_s^2/4\pi,\, \alpha_2 = g^2/4\pi,\, \alpha_1 = g'^2/4\pi$, respectively. ALP couplings $c_{\ff}$ to SM fermions are defined in the mass basis. We assume that fermion couplings are flavor-universal and flavor-diagonal at the cutoff scale, unless said otherwise. For quarks, this means
\begin{align}\label{eq:fu}
c_{\ff}(\Lambda) = c_{qq}(\Lambda),\qquad q = \{u,d,c,s,t,b\}.
\end{align}
This assumption implies that the ALP couples to fermions with a SM-like mass hierarchy. 

The renormalization group evolution of the couplings to lower scales $\mu < \Lambda$ is known at the leading logarithmic order~\cite{Choi:2017gpf,MartinCamalich:2020dfe,Chala:2020wvs,Bauer:2020jbp}. In this work, we combine observables at different scales, from several TeV down to the bottom mass. RG effects are highly relevant in this combination, because ALP couplings mix and evolve under the renormalization group. We have written a code called \href{https://github.com/TdAlps/TdAlps}{TdAlps} that implements the RG evolution and matching conditions from Ref.~\cite{Bauer:2020jbp} for ALP couplings at scales $\mu > 1\,$GeV.

After spontaneous symmetry breaking, ALP couplings to the photon $\gamma$ and the $Z$ and $W$ bosons in unitary gauge read
\begin{align}\label{eq:lagrangian-gauge}
    \mathcal{L}_{\rm eff} & \supset \frac{\alpha}{4\pi}\frac{a}{f_a}\left(c_{\gamma \gamma}\,F_{\mu\nu}\widetilde{F}^{\mu\nu} + 2\frac{c_{\gamma Z}}{s_w c_w}F_{\mu\nu}\widetilde{Z}^{\mu\nu} + \frac{c_{ZZ}}{s^2_w c^2_w}Z_{\mu\nu}\widetilde{Z}^{\mu\nu} + 2\frac{c_{WW}}{s^2_w}\,W^+_{\mu\nu} \widetilde{W}^{-\mu\nu}\right).
\end{align}
Here $\alpha$ is the electromagnetic coupling, $s_w = \sin\theta_w$ and $c_w = \cos\theta_w$ are the sine and cosine of the weak mixing angle, and $F_{\mu\nu},Z_{\mu\nu},W_{\mu\nu}$ are the field strength tensors of the physical gauge bosons. The couplings
\begin{align}\label{eq:ew-couplings}
c_{\gamma\gamma} = c_{WW} + c_{BB},\quad c_{\gamma Z} = c^2_w\, c_{WW} - s^2_w c_{BB},\quad c_{ZZ} = c^4_w \, c_{WW} + s^4_w\, c_{BB}
\end{align}
are defined at the weak scale $\mu_w = m_Z$. At energies above the weak scale, $c_{WW}$ and $c_{BB}$ are not renormalized~\cite{Bonilla:2021ufe}. This allows us to express the couplings in~\eqref{eq:ew-couplings} directly in terms of $c_{WW}$ and $c_{BB}$ at the cutoff scale $\Lambda$.

In the effective Lagrangian~\eqref{eq:lagrangian}, the ALP mass $m_a$ is a free parameter. For concreteness, we set
\begin{align}\label{eq:ma}
    m_a = 300\,\text{MeV}
\end{align}
throughout this work. The LHC and LEP observables we consider are insensitive to the benchmark choice for ALPs with masses $m_a \ll m_Z$. For flavor observables, the benchmark~\eqref{eq:ma} implies that ALPs can be resonantly produced in $B$ meson decays and decay mostly into muons. As a consequence, our interpretation of flavor data and the combination with LHC results are benchmark-dependent.

\begin{figure}[t!]
    \centering
    \includegraphics[width=0.45\textwidth]{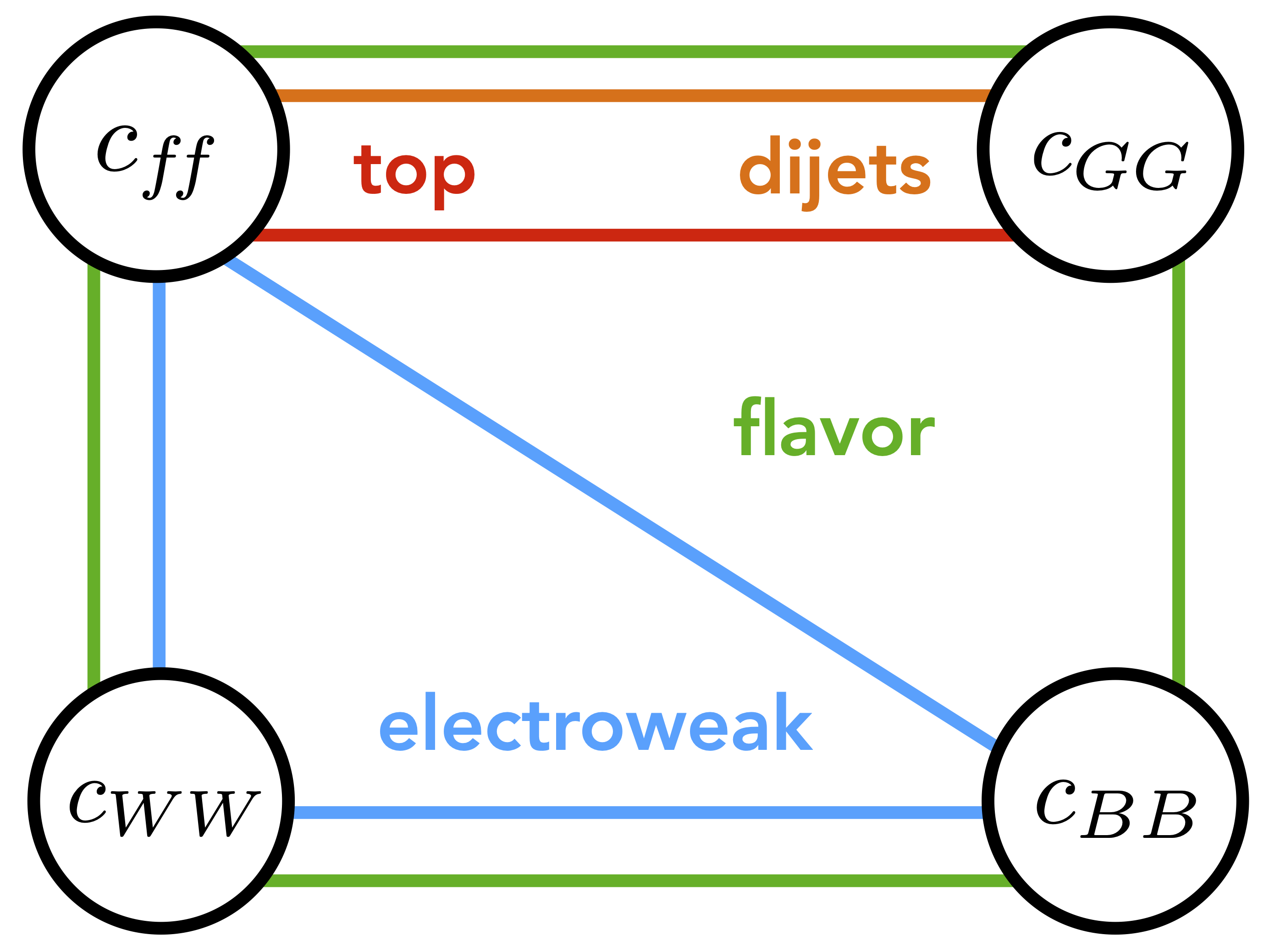}
     \caption{Overview of the observables used in our analysis. The colored lines illustrate which ALP couplings are mostly probed by each sector.\label{fig:alp-graphic}}
\end{figure}

The main goal of this work is to resolve the full parameter space of ALP couplings, $\{c_{\ff},\,c_{GG},\,c_{WW},\,c_{BB}\}$. We combine high-energy observables at the LHC and LEP, as well as flavor observables at LHCb and Belle II that constrain the parameter space in all directions. In Fig.~\ref{fig:alp-graphic}, we illustrate the sensitivity of the different sectors to the ALP couplings. Top, dijet and electroweak observables are sufficient to resolve the full parameter space. Flavor observables are sensitive to an intricate combination of all four parameters. When added to the high-energy observables, they drastically enhance the sensitivity of the global fit.

This analysis serves as a proof of principle and contains only a small selective set of ob\-ser\-va\-bles. The sensitivity of the global fit can be further improved in a more comprehensive data analysis.

\section{ALPs at the LHC}\label{sec:lhc}
At the LHC, ALPs can be probed through resonances or virtual effects in many different signatures. Virtual effects of light ALPs in high-energy observables depend neither on the mass, nor on the decay width or decay modes of the ALP. Searches for virtual ALPs are thus less model-dependent than searches for resonances, but typically also less sensitive. In this work, we prioritize resolving the ALP parameter space for a large class of models. We therefore focus on virtual ALP effects in high-energy processes,\footnote{In this paper, we refer to high-energy observables at ATLAS and CMS as \emph{LHC observables}. Flavor observables at LHCb will be discussed in Sec.~\ref{sec:flavor}.} keeping in mind that the sensitivity to the ALP couplings in specific scenarios could be enhanced by adding resonance searches.

To resolve the full ALP parameter space $\{c_{\ff},\,c_{GG},\,c_{WW},\,c_{BB}\}$ with LHC and LEP data, we strategically select sensitive observables. Top and dijet observables (Sec.~\ref{sec:top-dijet}) probe $c_{\ff}$ and $c_{GG}$, while effects of other couplings are sub-dominant. Electroweak observables (Sec.~\ref{sec:ew}) are sensitive to $c_{WW}$ and $c_{BB}$, but also depend on $c_{\ff}$. Our selection is motivated by the fact that the dominant effects in these observables involve as few ALP couplings as possible. This allows us to resolve the sub-space $\{c_{GG},c_{\ff}\}$ from top and dijet observables alone. ALP couplings to tops and gluons play a special role in the ALP effective theory, because they induce particularly large RG contributions to other ALP couplings at energies well below the cutoff scale~\cite{Bauer:2017ris,Choi:2017gpf,MartinCamalich:2020dfe,Chala:2020wvs,Bauer:2020jbp}. A self-consistent analysis of $c_{GG}$ and $c_{tt} = c_{\ff}$ in high-energy observables allows us to derive robust and model-independent bounds on effects of these couplings in low-energy observables.

Previous analyses of virtual ALPs in high-energy observables have focused on one or two ALP couplings. Bounds on the ALP-top coupling $c_{tt}$ have been obtained from top-antitop production, assuming that the ALP couples only to top quarks~\cite{Bonilla:2021ufe,Esser:2023fdo} or to tops and gluons~\cite{Galda:2021hbr}. Dijet distributions as a probe of $c_{GG}$ have been analyzed in Ref.~\cite{Gavela:2019cmq}. Observables in di-boson production have been used to set bounds on the individual couplings $c_{\gamma\gamma}$ and $c_{ZZ}$ for a fixed gluon coupling $c_{GG}$~\cite{Gavela:2019cmq,Carra:2021ycg}. Independent bounds on the weak gauge couplings $c_{BB}$ and $c_{WW}$ have been derived from vector boson scattering~\cite{Bonilla:2022pxu} and from electroweak precision tests~\cite{Bauer:2017ris}. 

For this analysis, we confine ourselves to a minimal set of observables: dijet angular distributions, top-antitop momentum distributions, the cross section of four-top production, same-sign $WW$ production, and the decay width of the $Z$ boson. In our predictions of each observable, we include contributions from all four ALP couplings $c_{\ff}$, $c_{GG}$, $c_{WW}$ and $c_{BB}$. This allows us to explore the full parameter space of the ALP effective Lagrangian, without making assumptions about the strength of individual couplings at a certain scale. It also enables us to relate ALP effects in observables at different energy scales through the renormalization group. Both aspects are essential for a global analysis of the ALP effective theory and for interpreting the results in terms of UV completions.

For our numerical predictions of observables in the ALP effective theory, we have created a {\tt UFO} model for the Feynman rules derived from the Lagrangian~\eqref{eq:lagrangian} using {\tt FeynRules}~\cite{Alloul:2013bka}. Based on this {\tt UFO} model, we simulate events from proton-proton collisions at $\sqrt{s} = 13\,$TeV using {\tt MadGraph5-aMC@NLO}~\cite{Alwall:2014hca}. We work at leading order (LO) in QCD and use the {\tt nnpdf31-NLO-as-0118} set of parton distribution functions (PDFs) by the NNPDF colla\-boration~\cite{NNPDF:2017mvq}, which does not use LHC top data when fitting the PDF parameters. Final states are analyzed at the parton level. Our treatment of the SM predictions and the corresponding theory uncertainties depends on the observable; we will discuss it case by case in the sections below.

Our parameter fits are performed using the {\tt SFitter} software~\cite{Lafaye:2004cn,Lafaye:2007vs,Lafaye:2009vr}. In the fits, we apply the theory uncertainties on the SM prediction of observables to the full prediction in the ALP effective theory. To take account of the fact that ALP and SM contributions may depend differently on scales and input parameters, we are conservative about the overall uncertainty.

\subsection{ALPs in top and dijet observables}
\label{sec:top-dijet}
In top and dijet production, the ALP enters as a virtual particle in the $s$ or $t$ channel,
\begin{align}
    pp \stackrel{a^\ast}{\longrightarrow} \{t\bar{t},\,jj\}\,.
\end{align}
At the leading order in QCD, ALP contributions to the amplitudes scale as
\begin{align}\label{eq:amp-lo}
\mathcal{M}^{(0)}(gg \stackrel{a^\ast}{\longrightarrow} t\bar{t}) & \sim \frac{\alpha_s}{4\pi}\,c^{\rm eff}_{GG}(\sqrt{s})\,c_{\ff}(\sqrt{s})\\\nonumber
\mathcal{M}^{(0)}(gg \stackrel{a^\ast}{\longrightarrow} gg) & \sim \left(\frac{\alpha_s}{4\pi}\, c^{\rm eff}_{GG}(\sqrt{s})\right)^2,
\end{align}
and $\sqrt{s}$ is the partonic center-of-mass energy. Amplitudes with external light quarks play no role; they are strongly suppressed by the small quark mass. The factor $\alpha_s/4\pi$ appears because of our assumption that the ALP-gluon coupling is loop-induced in UV completions of the effective theory, see~\eqref{eq:lagrangian}. The effective ALP-gluon coupling is defined as
\begin{align}\label{eq:anomaly-shift}
c^{\rm eff}_{GG}(\mu) & = c_{GG}(\mu) + \sum_{q} \frac{c_{qq}(\mu)}{2}\,\theta(\mu - m_q)\\\nonumber
& \stackrel{\rm FU}{=} c_{GG}(\mu) + \frac{c_{\ff}(\mu)}{2}\,N_q,\quad m_q < \mu.
\end{align}
In the first line, the sum includes all quarks with masses below the energy scale $\mu$ at which the coupling is probed. In the flavor-universal (FU) case, the second term in~\eqref{eq:anomaly-shift} counts the number of light quarks $N_q$ in the theory. This contribution of light quarks is not suppressed by the small quark mass and does not decouple at high energies. Any high-energy observable that involves the effective ALP-gluon coupling is sensitive to a linear combination of $c_{GG}$ and $c_{\ff}$ and thereby to the flavor structure of the ALP effective theory.

\begin{figure}[t!]
    \centering
    \includegraphics[width=0.35\textwidth]{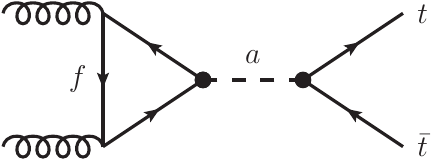}\hspace*{1.8cm}
     \includegraphics[width=0.29\textwidth]{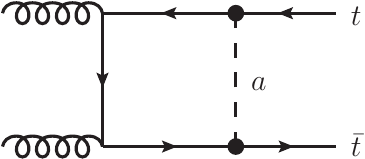}\vspace*{0.3cm}
     \caption{Feynman diagrams for ALP contributions to $t\bar{t}$ production. Left: contributions from the axial vector anomaly. In the loop, all quarks $f$ with masses below the momentum flowing through the vertex contribute to a shift of the ALP-gluon coupling. Right: chirality-suppressed contributions in QCD at NLO.}
    \label{fig:alps-nlo}
\end{figure}

\paragraph{Axial vector anomaly} The origin of light-quark contributions to the effective ALP-gluon coupling is the axial vector anomaly in QCD. In a theory with conserved vector currents like QCD, the divergence of the axial vector current $j_5^\mu$ for a single quark flavor $f$ reads
\begin{align}
\partial_{\mu} j_5^\mu = \frac{\alpha_s}{4\pi}\, G_{\mu\nu} \widetilde{G}^{\mu\nu},\qquad j_5^\mu = \bar{f} \gamma^\mu \gamma_5 f,
\end{align}
where the sum over gauge indices is implicit. Expressed in terms of the gluon fields $G_\mu$, this leads to\footnote{The trace $\text{Tr}[\dots]$ contracts the color charges of the gluon fields in a gauge singlet.}
\begin{align}
\partial_{\mu} j_5^\mu = \frac{\alpha_s}{4\pi} \epsilon^{\mu\nu\rho\sigma} \Big\{4\,\text{Tr}\big[\partial_\mu G_\nu \partial_\rho G_\sigma\big] + 4 ig_s\, \text{Tr}\big[ \partial_\mu G_\nu [G_\rho,G_\sigma] \big] \Big\}.
\end{align}
The two terms on the right side of the equation correspond to anomalies in triangle and box diagrams with two and three external gluons, respectively.\footnote{The anomalies for triangle, box and pentagon graphs have been calculated in a more general framework in Ref.~\cite{Bardeen:1969md}. Pentagon diagrams with four external gluons do not feature an axial anomaly in the ALP effective theory.} At the amplitude level, they induce a shift of the ALP-gluon coupling in the $agg$ and $aggg$ vertices as in~\eqref{eq:anomaly-shift}. An example of this effect in $t\bar{t}$ production is shown in Fig.~\ref{fig:alps-nlo}, left. A similar diagram can be drawn for dijet production by replacing the outgoing top quarks with gluons.

ALP couplings to the weak gauge bosons are not affected by an anomaly, because the $SU(2)_L$ gauge theory with SM fermions is anomaly-free.

\paragraph{Cross section at LO} The partonic cross section for $gg\stackrel{a^\ast}{\longrightarrow} \{t\bar{t},gg\}$ can be written as
\begin{align}\label{eq:cross-section}
\sigma(s) = \sigma_{\rm SM}(s) + 2\,\frac{c^{\rm eff}_{GG}(\sqrt{s})\,c_{X}(\sqrt{s})}{f_a^2}\,\bar{\sigma}_{a-\text{SM}}(s) + \frac{\left(c_{GG}^{\rm eff}(\sqrt{s})\,c_{X}(\sqrt{s})\right)^2}{f_a^4}\,\bar{\sigma}_{a-a}(s)\,,
\end{align}
where $c_X= \{c_{\ff},c_{GG}^{\rm eff}\}$. The partonic cross section in the Standard Model is denoted by $\sigma_{\rm SM}$; $\bar{\sigma}_{a-\text{SM}}$ and $\bar{\sigma}_{a-a}$ correspond to (normalized) contributions from ALP-SM and ALP-ALP amplitude interference.

A crucial difference between ALP effects in dijet and top-antitop production is their kinematic behavior at high energies. In dijet production, the three contributions to the cross section~\eqref{eq:cross-section} for $\sqrt{s} \gg m_a$ scale as~\cite{Baldenegro:2018hng,Gavela:2019cmq}
\begin{align}\label{eq:diboson-highenergy}
\sigma_{\text{SM}}(s) \sim \frac{1}{s}\,,\qquad \sigma_{a-\text{SM}}(s) \sim \frac{1}{s}\frac{s}{f_a^2}\,,\qquad \sigma_{a-a}(s) \sim \frac{1}{s}\frac{s^2}{f_a^4}\,.
\end{align}
Compared with the SM cross section, ALP contributions are enhanced at high energies. This is a typical feature in an effective theory, where scattering processes can diverge at energies around the cutoff scale and need to be regularized by new physics.

In top-antitop production, the energy enhancement of the ALP is tamed because the ALP-top coupling involves a chirality flip that scales as $m_t/\sqrt{s}$ in observables. At energies $\sqrt{s} \gg m_t, m_a$, the partonic cross sections scale as
\begin{align}\label{eq:ttb-highenergy}
\sigma_{\text{SM}}(s) \sim \frac{1}{s}\,,\qquad \sigma_{a-\text{SM}}(s) \sim \frac{1}{s}\frac{m_t^2}{f_a^2}\,,\qquad \sigma_{a-a}(s) \sim \frac{1}{s}\frac{m_t^2s}{f_a^4}\,.
\end{align}
In top observables, there is thus no energy enhancement in the interference of ALP and QCD amplitudes, only in the interference of two ALP amplitudes.

\paragraph{Beyond LO} At NLO in QCD, the amplitudes $\mathcal{M} = \mathcal{M}^{(0)} + \mathcal{M}^{(1)}$ receive ALP contributions scaling as
\begin{align}\label{eq:amp-nlo}
\mathcal{M}^{(1)}\big(gg \stackrel{a^\ast}{\longrightarrow} t\bar{t}\big) & \sim \frac{\alpha_s}{4\pi}\,c_{\ff}^2(\sqrt{s})\\\nonumber
\mathcal{M}^{(1)}\big(gg \stackrel{a^\ast}{\longrightarrow} gg\big) & \sim \left(\frac{\alpha_s}{4\pi} c^{\rm eff}_{GG}(\sqrt{s})\right)\frac{\alpha_s}{4\pi}\,c_{\ff}(\sqrt{s}).
\end{align}
Formally, these corrections are of the same order in $\alpha_s/4\pi$ as the LO amplitudes in~\eqref{eq:amp-lo}, due to the definition of the ALP-gluon coupling in~\eqref{eq:lagrangian}. From the point of view of perturbation theory, they should be taken into account in a consistent analysis of dijet and top-antitop production.

Including the full one-loop contributions, the $agg$ vertex function at NLO in QCD reads~\cite{Spira:1995rr,Bauer:2020jbp,Bonilla:2021ufe}
\begin{align}\label{eq:agg-full-nlo}
c_{GG}^{(1)}(\mu) = c_{GG}(\mu) + \sum_q \frac{c_{qq}(\mu)}{2}\, B_1\bigg(\frac{4 m_q^2}{\mu^2}\bigg), \qquad q = \{u,d,c,s,t,b\},
\end{align}
with the loop function
\begin{align}
B_1(x) = 1 - x\bigg[\frac{\pi}{2} + \frac{i}{2}\log\bigg(\frac{1 + \sqrt{1-x}}{1 - \sqrt{1-x}}\bigg)\bigg]^2,\qquad x < 1.
\end{align}
The first term in $B_1(x)$ is the contribution from the axial anomaly. The remainder tends to zero as $x\to 0$. At high energies $\sqrt{s} \gg 2 m_q$, the loop function in~\eqref{eq:agg-full-nlo} thus tends to 1 and only the anomaly contribution remains, cf.~\eqref{eq:anomaly-shift}. In our numerical predictions, we include the full NLO contribution~\eqref{eq:agg-full-nlo}.

In dijet production, vertex corrections are the only ALP contributions that scale as in~\eqref{eq:amp-nlo}. This implies that our predictions of dijet observables include all leading contributions of $c_{GG}$ and $c_{\ff}$, up to QCD corrections to LO amplitudes. In $t\bar{t}$ production, further NLO contributions arise from ALP insertions with two top coup\-lings in the LO SM amplitude for $gg\to t\bar{t}$. ALP contributions to the top-antitop-gluon coupling have been calculated in Ref.~\cite{Galda:2021hbr}. However, ALPs also induce virtual effects beyond vertex corrections, as for example in Fig.~\ref{fig:alps-nlo}, right, or can be radiated off the top quarks in the final state. A sound prediction of the dominant ALP effects in top-antitop production requires a complete calculation of all effects that scale as~\eqref{eq:amp-nlo}. Since this task exceeds the scope of this work, we work around it. At high energies, all contributions scaling as in~\eqref{eq:amp-nlo} are chirality-suppressed and decouple as $m_t/\sqrt{s} \to 0$, because the ALP couples to on-shell tops through a derivative. For our analysis, we select top observables at energies $\sqrt{s} \gg 2 m_t$ and neglect NLO contributions that are either chirality-suppressed or of higher order in $\alpha_s$. Our predictions do include the known $agg$ vertex correction from~\eqref{eq:agg-full-nlo}.

A third process that probes $c_{GG}$ and $c_{\ff}$ is four-top production. At the tree level, ALPs contribute through two kinds of amplitudes,
\begin{align}\label{eq:amp-4t}
\mathcal{M}_1^{(0)}\big(gg \stackrel{a^\ast}{\longrightarrow} t\bar{t}t\bar{t}\big) & \sim \left(\frac{\alpha_s}{4\pi}\,c_{GG}^{\rm eff}(\sqrt{s})\right) c_{\ff}(\sqrt{s}),\\\nonumber
\mathcal{M}_2^{(0)}\big(gg  \stackrel{a^\ast}{\longrightarrow} t\bar{t}t\bar{t}\big) & \sim c_{\ff}^2(\sqrt{s}).
\end{align}
In the first amplitude, the ALP is exchanged between gluons and tops, as in top-antitop production~\eqref{eq:amp-lo}. In the second amplitude, the ALP is exchanged between two top quarks, hence it is sensitive to the ALP-top coupling alone. Other contributions $\propto c^2_{\ff}$ through the axial anomaly or through NLO QCD corrections are suppressed by a relative factor of $\alpha_s/4\pi$. In flavor-universal scenarios, the leading ALP effects in $c_{\ff}$ are therefore well described by tree-level amplitudes.

Besides $agg$ couplings, also $aggg$ vertex corrections play a role in four-top production. In our analysis we include the anomaly contribution from~\eqref{eq:anomaly-shift} in both the $agg$ and $aggg$ couplings. A complete NLO analysis of four-top observables would improve the predictions in the ALP effective theory, but is beyond the scope of this work.

In what follows, we discuss the phenomenology of light ALPs in dijet, top-antitop and four-top production. As we will see, a combined fit of all three sectors is needed in order to constrain the $\{c_{\ff},c_{GG}\}$ parameter space without leaving blind directions.

\paragraph{Dijet production}
\label{sec:dijets}
At the LHC, dijet production probes strong interactions at energies up to several TeV. Here we use dijet angular distributions to constrain the effects of virtual ALPs described above. 

Our analysis is based on measurements of dijet angular distributions by the CMS collaboration, based on $35.9\,$fb$^{-1}$ of LHC data at 13 TeV~\cite{CMS:2018ucw}. Angular correlations between the two jets are measured in terms of rapidity differences,
\begin{align}
\frac{1}{\sigma(M_{jj})}\frac{d\sigma(M_{jj})}{d\chi_{jj}},\quad \chi_{jj} = \exp\big(|y_1 - y_2|\big),
\end{align}
where $y_1$ and $y_2$ are the rapidities of the two jets with the highest transverse momenta in an event. The distributions are reported in bins of the dijet invariant mass, $M_{jj}$, normalized to the cross section in the respective bin, $\sigma(M_{jj})$. Following~\eqref{eq:cross-section}, the prediction of the dijet differential distribution in the ALP effective theory reads
\begin{align}
\frac{d\sigma}{d\chi_{jj}} = \left(\frac{d\sigma}{d\chi_{jj}}\right)_{\rm SM} + \frac{\text{Re}\Big[\big(c_{GG}^{(1)}(\sqrt{s})\big)^2\Big]}{f_a^2} \left(\frac{d\bar{\sigma}}{d\chi_{jj}}\right)_{a-\text{SM}} + \frac{\Big| \big(c_{GG}^{(1)}(\sqrt{s})\big)^2\Big|^2}{f_a^4}\left(\frac{d\bar{\sigma}}{d\chi_{jj}}\right)_{a-a}.
\end{align}
For the SM prediction, we adopt the NLO QCD results from Ref.~\cite{CMS:2018ucw}. For the ALP contributions, we perform simulations at LO in QCD, as described at the beginning of Sec.~\ref{sec:lhc}. We include the full NLO vertex corrections from~\eqref{eq:agg-full-nlo} by rescaling the ALP-gluon couplings in the LO results.

\begin{figure}[t!]
    \centering
    \includegraphics[width=0.49\textwidth]{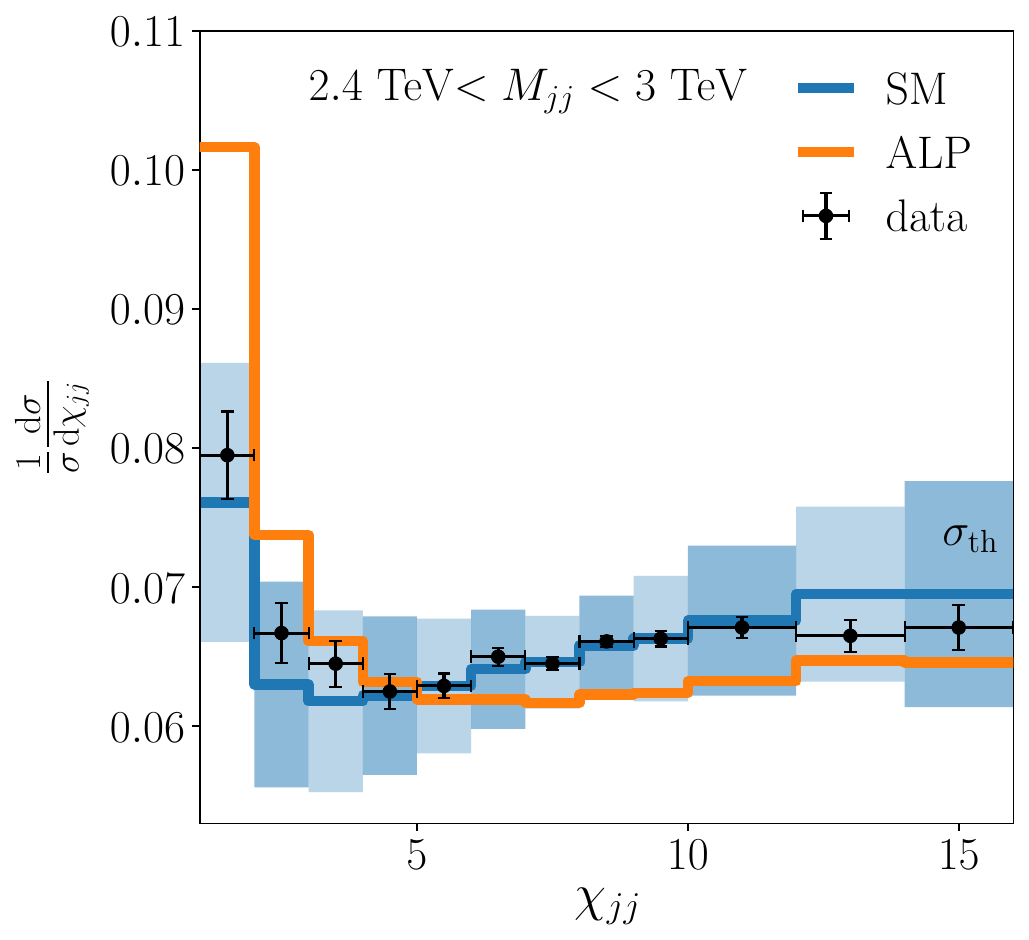}\hspace*{0.5cm}\raisebox{-0.3cm}{
     \includegraphics[width=0.466\textwidth]{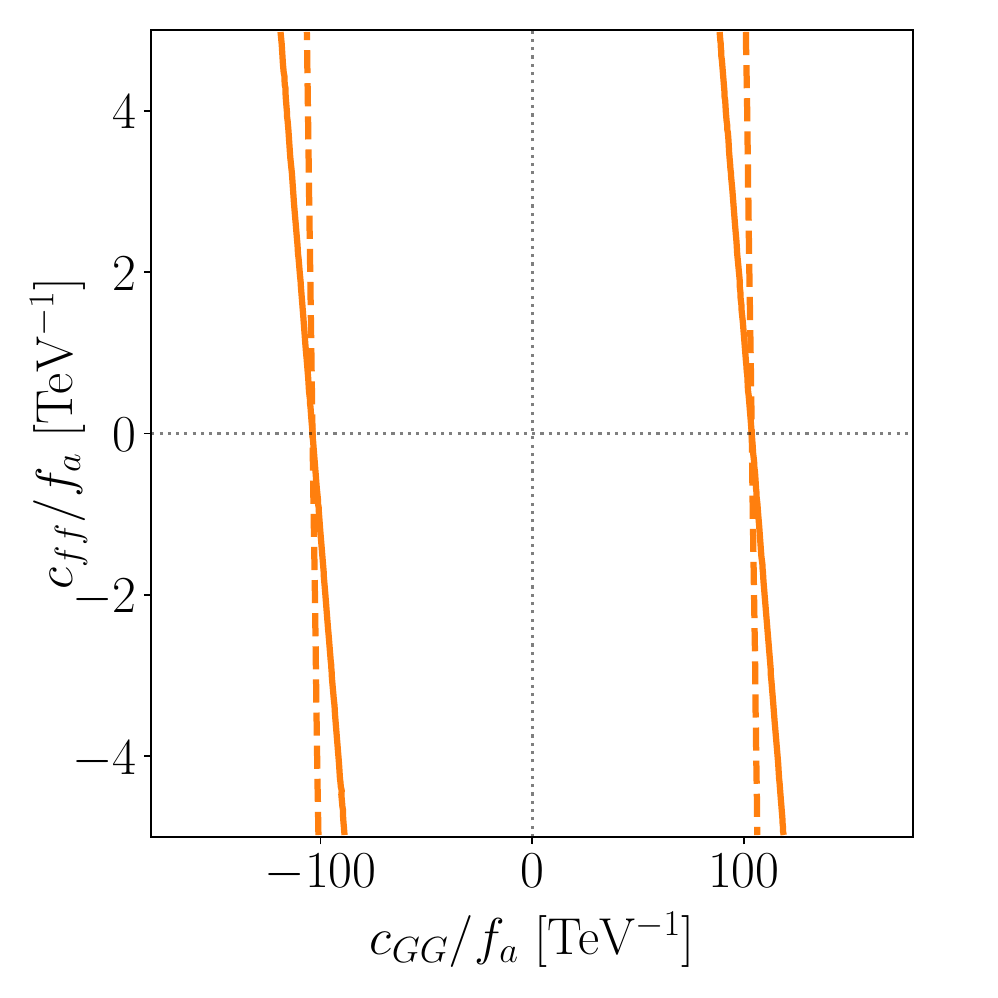}}
     \caption{\label{fig:dijet-angular}ALP effects in dijet production at the 13-TeV LHC. Left: dijet angular distribution in QCD (blue), in the ALP effective theory for $\alpha_s\,|c_{GG}^{\rm eff}|/(4\pi f_a) = 0.646/$TeV (orange), and as measured by the CMS collaboration (black)~\cite{CMS:2018ucw}. The theory uncertainties on the QCD prediction in each bin, $\sigma_{\rm th}$, are indicated as blue bars around the central value. Right: bounds on the $\{c_{\ff},c_{GG}\}$ parameter space from a combined fit to the darker shaded bins in the left panel. The contours correspond to $\Delta \chi^2 = 5.99$ in a 2-parameter fit. The region outside the two lines is excluded. Plain orange lines correspond to flavor-universal ALP-fermion couplings~\eqref{eq:fu}; dashed lines indicate a scenario where the ALP couples only to top quarks and gluons.}
\end{figure}

In Fig.~\ref{fig:dijet-angular}, left, we show the dijet angular distribution for $2.4\,\text{TeV} < M_{jj} < 3\,\text{TeV}$, the lowest $M_{jj}$ bin in the CMS analysis. We choose this bin because it comprises the largest data set. Furthermore, the energies probed in this range lie below the cutoff scale $\Lambda = 4\pi\,$TeV, where we expect the ALP effective theory to be valid. 

For forward jet emission, $\chi \to \infty$, the QCD prediction is dominated by Rutherford scatte\-ring. By choosing $\chi_{jj}$ as a measure of dijet angular correlations, the forward singularity in QCD is subtracted from the cross section $d\sigma/d\chi_{jj}$, resulting in a rather flat distribution (in blue). In the forward region $\chi \to \infty$, ALP contributions resemble QCD, due to the exchange of a light virtual particle in the $t$-channel. For central jet emission $\chi_{jj} \to 1$, ALP contributions (orange) are enhanced compared to the QCD prediction. This is due to the different amplitudes obtained from pseudo-scalar (ALP) versus vector (gluon) exchange between the external gluons. At high energies, the effect is amplified by the energy scaling of the ALP contribution close to the cutoff scale, see~\eqref{eq:diboson-highenergy}.

For the parameter fit, we select only every second bin in the $\chi_{jj}$ distribution, in order to minimize (partly unknown) correlations between neighboring bins. The selected bins are shaded darker in Fig.~\ref{fig:dijet-angular}, left. We treat these bins as uncorrelated in the fit. For the theory uncertainty of the predictions in the ALP effective theory, we use the SM uncertainties from Ref.~\cite{CMS:2018ucw} for each bin and add an overall uncertainty of $5\%$ to be conservative. The orange curve in Fig.~\ref{fig:dijet-angular}, left, corresponds to $\Delta \chi^2 = 3.84$ in a 1-parameter fit of $c_{GG}^{\rm eff}/f_a$ to the set of dark shaded bins.

The effective ALP-gluon coupling $c_{GG}^{\rm eff}$ probes a linear combination of $c_{GG}$ and $c_{\ff}$, which is sensitive to the flavor structure of the ALP couplings to quarks, see~\eqref{eq:anomaly-shift}. In Fig.~\ref{fig:dijet-angular}, right, we show the results of a 2-parameter fit of $c_{GG}$ and $c_{\ff}$ to the dijet angular distributions from Fig.~\ref{fig:dijet-angular}, left. We distinguish between two flavor scenarios: The plain contours correspond to the flavor-universal scenario~\eqref{eq:fu}, where $c_{GG}^{\rm eff} = c_{GG} + 6\,c_{\ff}/2$. The dashed contours are obtained in a scenario where the ALP couples only to top quarks. In this case, $c_{GG}^{\rm eff} = c_{GG} + c_{\ff}/2$ and dijet production is much less sensitive to the ALP-fermion coupling.

\paragraph{Top-antitop production}
\label{sec:ttbar}
Precision measurements of top-antitop observables constrain the ALP coupling to top quarks and gluons. In the ALP effective theory, the structure of the $t\bar{t}$ cross section is given by~\eqref{eq:cross-section}. At the leading order, $t\bar{t}$ production is only sensitive to the product of couplings $c^{\rm eff}_{GG}\cdot c_{\ff}/f_a^2$, see~\eqref{eq:amp-lo}. NLO contributions change the probed direction in the $\{c_{\ff},c_{GG}\}$ space, see~\eqref{eq:amp-nlo}. 

\begin{figure}[t!]
    \centering
    \includegraphics[width=0.49\textwidth]{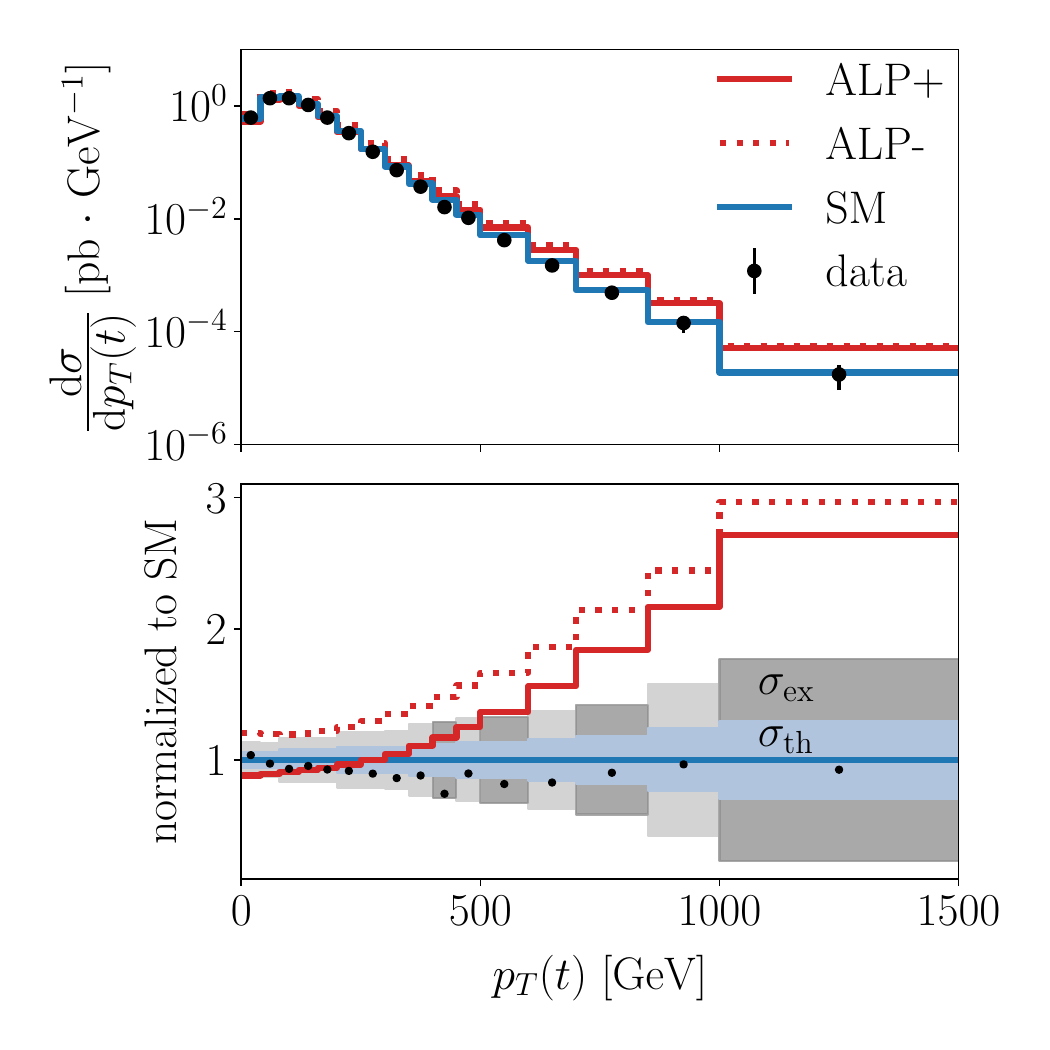}\hspace*{0.2cm}
     \raisebox{0.12cm}{\includegraphics[width=0.475\textwidth]{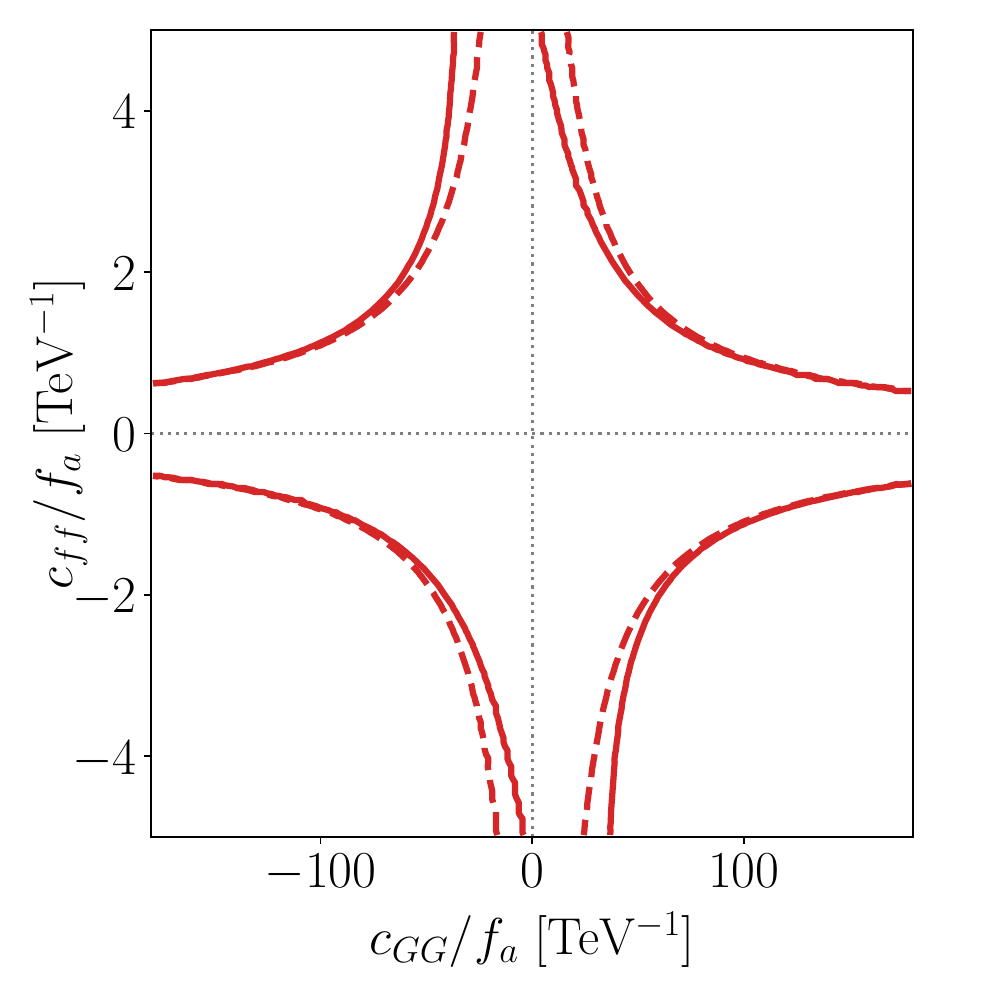}}
     \caption{ALP effects in top-antitop production at the 13-TeV LHC. Upper left: distribution of the top's transverse momentum, $p_T(t)$, in the SM (blue), in the ALP effective theory (red) for $\alpha_s c_{GG}^{\rm eff}c_{\ff}/(4\pi f_a^2) = \pm 1/$TeV$^{2}$, and as measured by CMS~\cite{CMS:2021vhb} (black data points). Lower left: the same distributions normalized to the SM prediction, including theory uncertainties $\sigma_{\rm th}$ (blue band) and stacked experimental uncertainties $\sigma_{\rm ex}$ (grey areas). Right: bounds on the $\{c_{\ff},c_{GG}\}$ parameter space from a combined fit to the dark shaded bins in the left panel. The contours correspond to $\Delta \chi^2 = 5.99$ in a 2-parameter fit. The region outside the curves is excluded. Plain red curves correspond to flavor-universal ALP-fermion couplings~\eqref{eq:fu}; dashed curves indicate a scenario where the ALP couples only to top quarks and gluons.\label{fig:top}}
\end{figure}

To derive bounds on the parameter space, we choose a measurement of the $t\bar{t}$ cross section at 13 TeV in bins of the top's transverse momentum, $p_T(t)$~\cite{CMS:2021vhb}. The CMS collaboration reports the differential distribution
\begin{align}\label{eq:toppt}
\frac{d\sigma}{d p_T(t)},
\end{align}
reconstructed from lepton+jets final states. In Fig.~\ref{fig:top}, left, the upper panel shows the CMS measurement (black points) compared to predictions in the Standard Model (blue) and in the ALP effective theory (red) for $\alpha_s c_{GG}^{\rm eff}c_{\ff}/(4\pi f_a^2) = \pm 1/$TeV$^{2}$. For the SM prediction, we use the results quoted in Ref.~\cite{CMS:2021vhb}. For the ALP contributions, we use our LO QCD simulations and include the vertex corrections from the axial anomaly according to~\eqref{eq:anomaly-shift}. At high $p_T(t)$, we expect this prediction to be a good approximation of the full NLO result, as discussed in Sec.~\ref{sec:top-dijet}.

In the lower panel of Fig.~\ref{fig:top}, left, we show the individual contributions to the $p_T(t)$ distribution normalized to the SM prediction. For both SM and ALP contributions, we assign a conservative theory uncertainty of twice the values quoted for the Standard Model in Ref.~\cite{CMS:2021vhb} (blue band). The ALP interference with the QCD amplitude is negative, leading to a higher event rate for $c_{GG}^{\rm eff}c_{\ff} < 0$ (dotted red) than for $c_{GG}^{\rm eff}c_{\ff} > 0$ (plain red). The relative enhancement of the ALP contribution at high momenta, see~\eqref{eq:ttb-highenergy}, is clearly visible.

For the parameter fit, we use the bins number 10, 12, 14 and 16 in the $p_T(t)$ distribution (shaded dark grey). In this way, we reduce correlations between adjacent bins. By focusing on high transverse momenta, we maximize the sensitivity to the ALP contribution and ensure that our approximate NLO prediction is applicable. The results of our 2-parameter fit to $\{c_{GG},c_{\ff}\}$ are shown in Fig.~\ref{fig:top}, right. Overall, the bounds on the ALP couplings are significantly stronger than those obtained from dijet distributions, cf. Fig.~\ref{fig:dijet-angular}. The blind direction along $|c_{GG}^{\rm eff}\cdot c_{\ff}| = 0$ is however apparent. As for the dijet distributions, we show the results for the flavor-universal scenario (plain) and the top-only scenario (dashed), which probe distinguishable regions of the ALP parameter space. Including the total $t\bar{t}$ cross section in the fit or using the normalized $p_T(t)$ distribution has no significant effect on these parameter bounds.

\paragraph{Four-top production}
\label{sec:4tops}
In our analysis of dijet and top-antitop production, we have encountered blind directions in the $\{c_{\ff},c_{GG}\}$ parameter space along
\begin{align}
c_{GG}^{\rm eff} = 0\ \ \text{(dijets)}  \qquad \text{and} \qquad |c_{GG}^{\rm eff}\cdot c_{\ff}| = 0\ \ \text{($t\bar{t}$)},
\end{align}
sub-leading NLO corrections left aside. Four-top production at the LHC resolves both of these directions through the tree-level contribution $\mathcal{M}^{(0)}\propto c_{\ff}^2$, see~\eqref{eq:amp-4t}.

The parameter directions probed by four-top production are hard to guess, because ALPs can contribute through many different Feynman diagrams. Moreover, multiple insertions of ALP propagators in an amplitude lead to effects that are comparable in size with single ALP insertions for $|c_{X}|/f_a \approx 1/$TeV. We include up to two ALP insertions in the amplitude. The cross section is then a polynomial of 8th order in $c_{GG}^{\rm eff}$ and $c_{\ff}$. At tree level, we obtain the total cross section for four-top production at $\sqrt{s} = 13\,$TeV as~\footnote{Here we neglect contributions with coefficients smaller than $0.1$ and round all contributions to two decimals. In the numerical analysis we use the full expression.}
\begin{align}\label{eq:xs-4t}
    \sigma_{4t} = \sigma_{\rm SM} & + \left[0.74\left(\frac{\alpha_s}{4\pi}c_{GG}^{\rm eff}\right)^2 + 2.45\,\frac{\alpha_s}{4\pi}c_{GG}^{\rm eff}c_{\ff} + 0.42\,c_{\ff}^2\right]\left(\frac{\rm TeV}{f_a}\right)^2 \text{fb}\\\nonumber
     & + \bigg[8.38\left(\frac{\alpha_s}{4\pi}c_{GG}^{\rm eff}\right)^4 + 3.73\left(\frac{\alpha_s}{4\pi}c_{GG}^{\rm eff}\right)^3c_{\ff}\\\nonumber
     & \qquad\ + 4.14\left(\frac{\alpha_s}{4\pi}c_{GG}^{\rm eff}\right)^2c_{\ff}^2 +  0.62\,\frac{\alpha_s}{4\pi}c_{GG}^{\rm eff}c_{\ff}^3\bigg]\left(\frac{\rm TeV}{f_a}\right)^4\text{fb}\\\nonumber
                & + \bigg[0.19\left(\frac{\alpha_s}{4\pi}c_{GG}^{\rm eff}\right)^4c_{\ff}^2 + 1.00\left(\frac{\alpha_s}{4\pi}c_{GG}^{\rm eff}\right)^3c_{\ff}^3\bigg]\left(\frac{\rm TeV}{f_a}\right)^6\text{fb}\\\nonumber
                & + 0.96\left(\frac{\alpha_s}{4\pi}c_{GG}^{\rm eff}\right)^4 c_{\ff}^4 \left(\frac{\rm TeV}{f_a}\right)^8\text{fb}.
\end{align}
For the SM prediction, we adopt the NLO prediction $\sigma_{\rm SM} = 11.97^{+18\%}_{-21\%}\,$fb from Ref.~\cite{Frederix:2017wme}, which includes both QCD and electroweak contributions. For the ALP contributions, we use our LO QCD simulations. To find the various contributions to the polynomial in~\eqref{eq:xs-4t}, we have performed several simulations, fixing either $c_{GG}^{\rm eff}$ or $c_{\ff}$ and varying the other parameter.

Electroweak contributions to four-top production with ALP insertions would introduce the couplings $c_{WW}$ and $c_{BB}$. However, at a similar coupling strength they are suppressed compared to the QCD-based contributions, due to the small quark luminosity from the colliding protons. Mixed QCD-electroweak contributions could be numerically relevant in ALP-SM interference, but do not introduce ALP couplings beyond $c_{GG}$ and $c_{\ff}$. We leave an exploration of these contributions for future work.

The four-top production cross section has been measured at the LHC by CMS~\cite{CMS:2019jsc,CMS:2019rvj,CMS:2023ica} and ATLAS~\cite{ATLAS:2020hpj,ATLAS:2021kqb,ATLAS:2023ajo} in different final states. The central values of the results from the two collaborations agree within uncertainties. In our fit, we include the results of all measurements, assuming the corresponding uncertainties to be uncorrelated.

\subsection{Combined fit to top and dijet data}
\label{sec:tg-fit}
\begin{figure}[t!]
    \centering
        \includegraphics[width=0.49\textwidth]{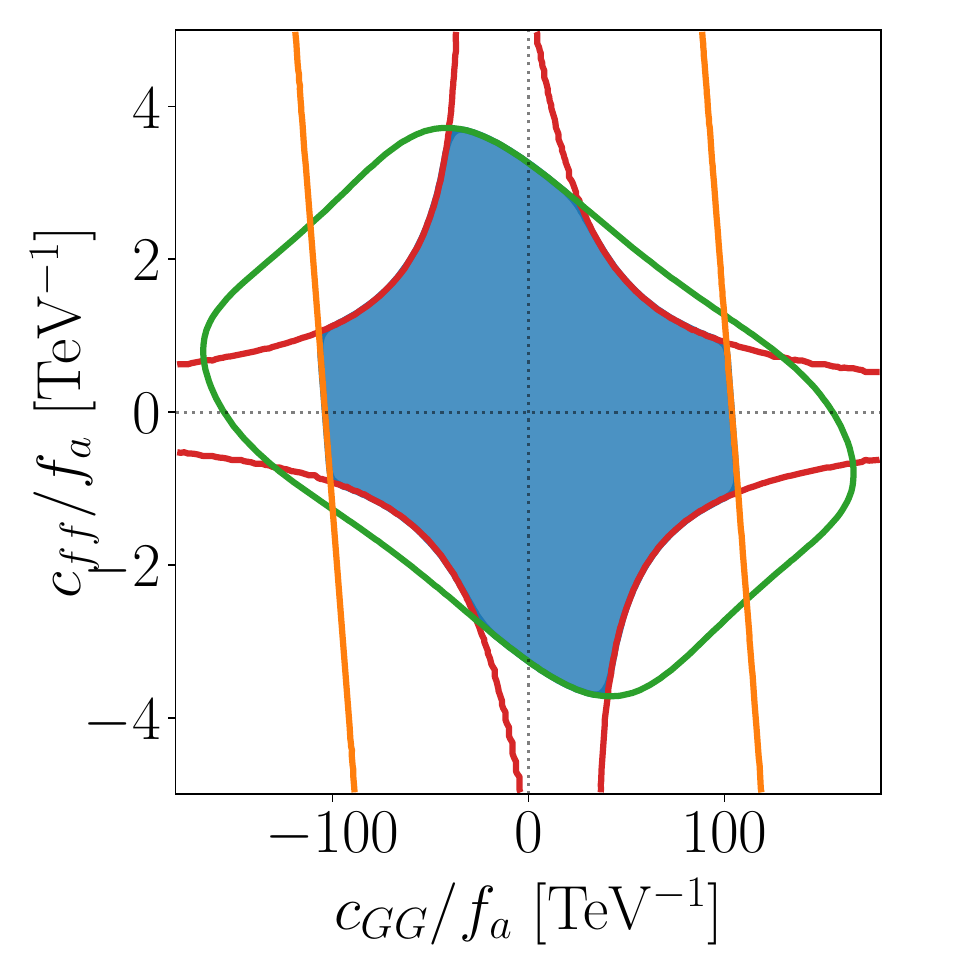}\hspace*{0.3cm}
        \includegraphics[width=0.49\textwidth]{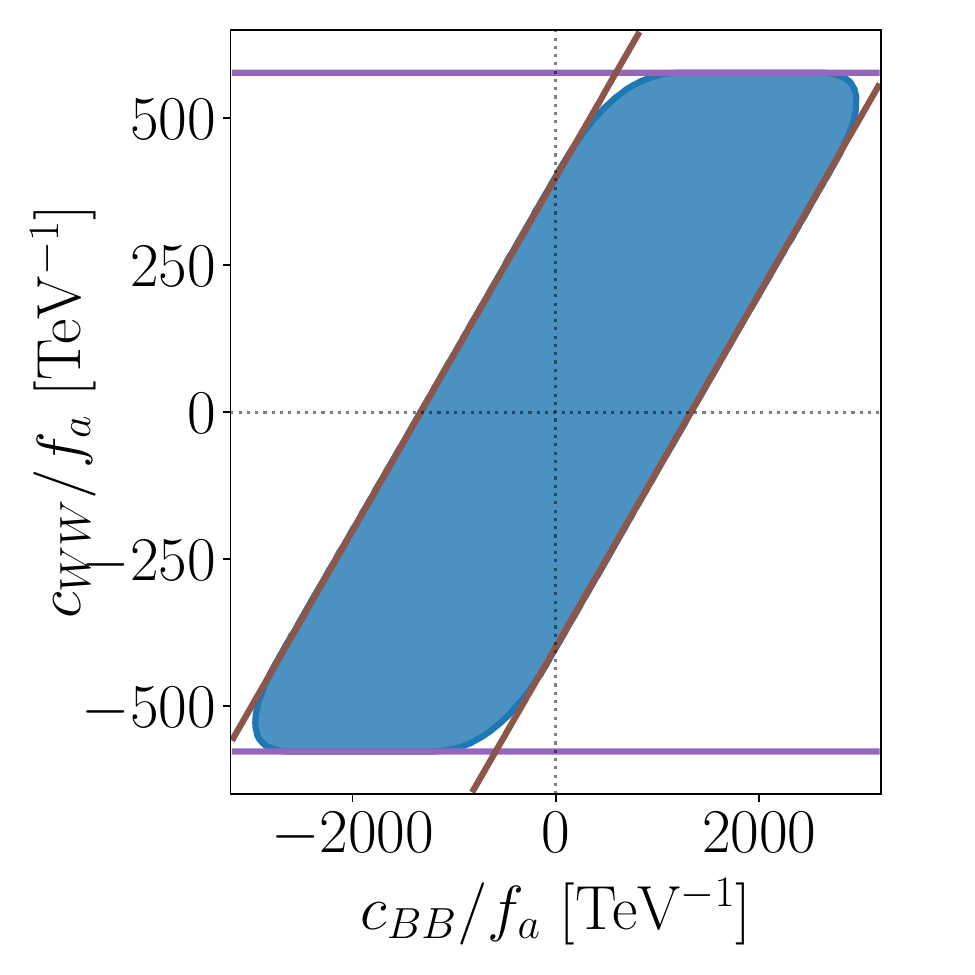}
     \caption{Bounds on the ALP parameter space with flavor-universal fermion couplings from measurements at the LHC and LEP. Left: individual 2-parameter fits of $\{c_{\ff},c_{GG}\}$ to measurements of dijet angular distributions (orange), top-antitop production at high transverse momenta (red), the four-top production cross section (green); and the result of the final  combined 4-parameter fit to all data from Secs.~\ref{sec:top-dijet} and \ref{sec:ew}, profiled over $c_{WW}$ and $c_{BB}$ (blue area). Right: 4-parameter fit of $\{c_{\ff},c_{GG},c_{WW},c_{BB}\}$ to all data from Secs.~\ref{sec:top-dijet} and \ref{sec:ew} (blue area), excluding same-sign diboson production (brown lines) or the $Z$ boson width (purple lines), and profiled over $c_{\ff}$ and $c_{GG}$.  The contours correspond to $\Delta \chi^2 = 5.99$ in the 2-parameter and (profiled) 4-parameter fits. The region outside the contours is excluded. All parameters $c_X$ are defined at the cutoff scale $\Lambda = 4\pi\,$TeV.\label{fig:global}}
\end{figure}

The fit results for top and dijet ob\-ser\-va\-bles are shown in Fig.~\ref{fig:global}, left, assuming flavor-universal ALP couplings to quarks as in~\eqref{eq:fu}. The colored contours correspond to bounds obtained from individual fits of $\{c_{\ff},c_{GG}\}$ to dijet angular distributions (orange), the transverse momentum distribution in top-antitop production (red), and the four-top production cross section (green). When combining the observables from all three sectors in one fit, we obtain the bounds corresponding to the blue area. It is apparent that four-top production constrains the blind directions left by dijet and top-antitop obser\-va\-bles. In the combined fit, the $\{c_{\ff},c_{GG}\}$ sector of the ALP parameter space is fully resolved.

For flavor-universal ALP-fermion couplings, the combined fit yields a robust upper bound of (see Fig.~\ref{fig:global} left)
\begin{align}\label{eq:cff-bound}
\frac{|c_{\ff}(4\pi\,\text{TeV})|}{f_a} \lesssim \frac{3.6}{\rm TeV}.
\end{align}
In scenarios without flavor universality, the correlation between $c_{\ff}$ and $c_{GG}$ in observables changes due to the anomaly contributions of light quarks to $c_{GG}^{\rm eff}$ from~\eqref{eq:anomaly-shift}. For the scenario with ALP-fermion couplings only to tops, the contours are distorted as shown in Figs.~\ref{fig:dijet-angular} and~\ref{fig:top}. Including four-top production adds little information on the flavor structure, because the ALP effects on the four-top cross section are dominated by $c_{\ff}$ in the relevant parameter region, see~\eqref{eq:xs-4t}. The bound on $c_{\ff}$ from~\eqref{eq:cff-bound} does therefore apply for both scenarios. It is still interesting to notice that a combined fit of $\{c_{\ff},c_{GG}\}$ to observables with and without external tops gives indirect information about ALP couplings to light quarks, which are typically hard to probe directly due to the quark mass suppression in observables.

The bound on $c_{\ff}$ from the combined top-dijet fit limits the contribution of virtual tops in effective ALP couplings at lower energies, see e.g. Refs.~\cite{Bauer:2017ris,Bauer:2021mvw,Bonilla:2021ufe}. Within a specific flavor scenario, this allows to extract independent information on the ALP couplings $c_{\ff}$, $f\neq t$, and $c_{VV}$ from low-energy observables.

\subsection{ALPs in electroweak observables}\label{sec:ew}
Electroweak observables probe ALP couplings to weak and hypercharge bosons, $c_{WW}$ and $c_{BB}$, at tree level. In $Z$-pole observables at LEP, these parameters determine the ALP couplings to $Z$ bosons and photons in linear combinations~\eqref{eq:ew-couplings}. However, the effective couplings $c_{\gamma Z}$ and $c_{\gamma\gamma}$ also involve the fermion couplings $c_{\ff}$ through loop contributions. Observables with external $W$ bosons give direct access to $c_{WW}$. But they are often also sensitive to $c_{GG}$, as for instance in $pp\to W^+W^-$ at the LHC. An independent analysis of $c_{WW}$ and $c_{BB}$ is therefore in general not possible. However, when combined with top and dijet observables, electroweak observables resolve the remaining two directions in the ALP parameter space that were left unconstrained by the QCD-based observables in Sec.~\ref{sec:top-dijet}.

For our analysis, we select two observables with a high sensitivity to the weak and hypercharge couplings: the total width of the $Z$ boson; and the cross section for same-sign $WW$ production at the LHC.

\paragraph{$Z$ boson width}\label{sec:z-width}
ALPs with masses below the $Z$ mass can induce exotic $Z$ decays. The partial decay width for $Z\to a\gamma$ is~\cite{Bauer:2017ris}
\begin{align}
    \Gamma(Z\to a\gamma) = \frac{\alpha \alpha(m_Z)}{96\pi^3 s_w^2 c_w^2}\frac{m_Z^3}{f_a^2} \left(c_{\gamma Z}^{(1)}(m_Z)\right)^2\left(1-\frac{m_a^2}{m_Z^2}\right)^3.
\end{align}
At the one-loop level, the ALP-$Z$-photon vertex receives contributions from fermion loops; electroweak corrections play a minor role~\cite{Bauer:2017ris}. The effective coupling is approximated as
\begin{align}\label{eq:alp-gamma-z}
    c_{\gamma Z}^{(1)}(m_Z) \approx c_{\gamma Z} + \sum_{f\neq t} N_c^f Q_f v_f \,c_{\ff}(m_Z),
\end{align}
where $v_f = T_3^f/2 - s_w^2 Q_f$ is the vector coupling of fermion $f$ to the $Z$ boson. The second term in~\eqref{eq:alp-gamma-z} is related to the axial anomaly and our choice of operator basis in the effective Lagrangian~\eqref{eq:lagrangian}. In this basis, the $Z$ width probes a linear combination of $c_{WW}$, $c_{BB}$ and $c_{\ff}$.

The precise measurement of the $Z$ boson width at LEP sets a bound on exotic decay modes. Comparing the measurement~\cite{ALEPH:2005ab} with the SM prediction~\cite{ParticleDataGroup:2020ssz}
\begin{align}
\Gamma_Z^{\rm exp} = 2.4952 \pm 0.0023\,\text{GeV},\qquad
 \Gamma_Z^{\rm SM} = 2.4942 \pm 0.0009\,\text{GeV}
\end{align}
constrains the branching ratio for $Z\to a\gamma$.

\paragraph{Same-sign $W^\pm W^\pm$ production}
To fully resolve the ALP parameter space, we need at least one more measurement that breaks the blind direction along $c_{\gamma Z}^{(1)} = 0$ in $\Gamma(Z\to a\gamma)$. Di-boson production at the LHC, $pp \to VV$, probes electroweak ALP couplings at tree level, but also involves the gluon coupling $c_{GG}$~\cite{Gavela:2019cmq,Carra:2021ycg}. Di-boson production in association with two jets, $pp\to VVjj$, is sensitive to vector boson scattering at high energies and probes $c_{WW}$ and $c_{BB}$ largely independently from $c_{GG}$~\cite{Bonilla:2022pxu}. Here we use same-sign $WW$ production, $pp\to W^\pm W^\pm jj$, via vector boson scattering to probe $c_{WW}$. This process does not involve $c_{GG}$ at tree level; it is sensitive to $c_{WW}$ alone.

Same-sign $WW$ production at the LHC has been measured in the leptonic final state as a function of the di-boson transverse mass $M_T(WW)$~\cite{CMS:2020gfh}. To obtain a bound on $c_{WW}$, we have implemented the analysis of Ref.~\cite{Bonilla:2022pxu}. The sensitivity to $c_{WW}$ is enhanced at high energies and thus in the tail of the $M_T(WW)$ spectrum. For this reason we only implement the last bin of the distribution.

\subsection{Combined fit to high-energy data}\label{sec:comb-lhc}
To constrain the $c_{WW}$ and $c_{BB}$ directions in the ALP parameter space, we perform a combined fit of all four parameters $\{c_{\ff},c_{GG},c_{WW},c_{BB}\}$ to the top, dijet and electroweak observables from Secs.~\ref{sec:top-dijet} and~\ref{sec:ew}. Such a global fit is necessary to resolve all directions in the parameter space, in particular the blind direction among $c_{WW}$, $c_{BB}$ and $c_{\ff}$ in $c_{\gamma Z}^{(1)}$~\eqref{eq:alp-gamma-z}.

In Fig.~\ref{fig:global}, right, we show the results of our 4-parameter fit to LHC and LEP data, including (blue) and excluding (brown) same-sign $WW$ production. For comparison, we also display the bound obtained by excluding the $Z$ width measurement (purple) and hence not constraining $c_{BB}$. As expected, the $\{c_{BB},c_{WW}\}$ parameter space is fully resolved. The bounds are dominated by the $Z$ width and same-sign $WW$ production, but implicitly involve top and dijet observables, which constrain the impact of $c_{\ff}$ on $\Gamma_Z$.

The sensitivity to $c_{BB}$ is lower than for $c_{WW}$, and the overall sensitivity to electroweak ALP couplings is lower than for ALP-gluon couplings, see Fig.~\ref{fig:global}, left. Loop effects of ALPs in electroweak precision tests~\cite{Bauer:2017ris,Aiko:2023trb} lead to somewhat stronger bounds on $c_{WW}$ and $c_{BB}$. Including these observables in our fit would further constrain the $\{c_{WW},c_{BB}\}$ parameter space. 

With our selection of observables, the sectors $\{c_{\ff},c_{GG}\}$ and $\{c_{BB},c_{WW}\}$ are only connected through the $Z$ width, which involves $c_{BB}$, $c_{WW}$ and $c_{\ff}$. Including further LHC observables would introduce more parameter correlations and enhance the sensitivity to the various directions in the parameter space. Besides the observables with virtual ALPs mentioned at the beginning of Sec.~\ref{sec:lhc}, resonance searches and observables with displaced vertices or missing energy can provide relevant information on the ALP couplings, see e.g.~\cite{Mimasu:2014nea,Bauer:2017ris,Esser:2023fdo,Rygaard:2023vyo}. Many of them, however, depend on the ALP's mass and decay width and/or are subject to additional model assumptions. Including them in a global analysis limits the generality of the fit results.

\section{ALPs at flavor experiments}
\label{sec:flavor}
Observables of flavor-changing neutral currents are sensitive probes of ALPs. Sub-GeV ALPs can be resonantly produced in two body decays $b\to s a$, $b\to d a$ and $s\to d a$ and probed in $B$ meson or kaon decays. Here we demonstrate the impact of flavor observables in a global analysis of the ALP effective Lagrangian. We focus on $B\to Ka,\, a\to \mu^+\mu^-$ decays, which can be significant relative to the rare $B\to K \mu^+\mu^-$ decays in the Standard Model. Depending on the ALP's lifetime, these decays can leave signatures of prompt, displaced or invisible muon pairs in an experiment. We will see that the lifetime plays a crucial role in determining which regions in the parameter space can be probed. 

The total event rate for $B\to K a,\, a\to \mu^+\mu^-$ decays is given by
\begin{align}
N_{\rm tot} =  N_B\,\mathcal{B}(B\to K a)\,\mathcal{B}(a \to \mu^+\mu^-),\qquad \mathcal{B}(a \to \mu^+\mu^-) =  \frac{\Gamma(a\to \mu^+\mu^-)}{\Gamma_a},
\end{align}
where $N_B$ is the number of produced $B$ mesons, $\mathcal{B}(x\to yz)$ is the branching ratio for the process $x\to yz$, and $\Gamma(a\to \mu^+\mu^-)$ and $\Gamma_a$ are the partial and total decay width of the ALP.

ALPs with flavor-diagonal couplings are produced in $B\to Ka$ decays through loops with virtual top quarks and $W$ bosons, relying mainly on $c_{tt}$ and $c_{WW}$. The total decay width $\Gamma_a$ depends on all couplings that enter the decay modes of the ALP. It normalizes the branching ratios of the ALP and sets its decay length. For an ALP with flavor-universal couplings to fermions, the event rate $N_{\rm tot}$ is thus a function of all four couplings $\{c_{\ff},\,c_{GG},\,c_{WW},\,c_{BB}\}$. Determining which directions in the ALP parameter space are probed by $B\to K a$ observables is more involved than for the LHC observables from Sec.~\ref{sec:lhc}, where a cross section scales with the relevant ALP couplings as in~\eqref{eq:cross-section}. We therefore briefly review the theory of ALPs in $B$ decays in Sec.~\ref{sec:alps-b-decays}. We then discuss the impact of the lifetime in Sec.~\ref{sec:flavor-signals} and analyze the impact of $B\to K a$ observables on the global fit of the ALP effective theory in Sec.~\ref{sec:flavor-impact}. 

\subsection{ALPs in $B\to K$ decays}
\label{sec:alps-b-decays}
At the quark level, the decay $b\to s a$ is induced by an effective ALP coupling to bottom and strange quarks. Below the weak scale $\mu_w$, the interaction is described by the effective Lagrangian
\begin{align}
    \mathcal{L}_{\rm eff}(\mu < \mu_w) = C_{sb}(\mu)\,\frac{\partial^{\mu} a}{f_a} (\bar{s}_L \gamma_{\mu} b_L) + h.c.
\end{align}
The Wilson coefficient $C_{sb}(\mu)$ is generated through matching contributions of loops with top quarks and $W$ bosons. Using results from~\cite{Bauer:2020jbp} and the \href{https://github.com/TdAlps/TdAlps}{TdAlps} code to RG-evolve the ALP parameters, we obtain the flavor-changing Wilson coefficient at the weak scale as
\begin{equation}\label{eq:csb-coupling}
C_{sb}(\mu_w) = \left(7.5499\,c_{\ff} - 0.0224\,c_{GG} - 0.0119\,c_{WW} - 0.0001\,c_{BB}\right)\cdot 10^{-4},
\end{equation}
where all ALP couplings are defined at the cutoff scale $\Lambda = 4\pi\,$TeV and flavor universality is assumed.
Below the weak scale, the RG evolution of the coupling is mild, so that $C_{sb}(\mu < \mu_w) = C_{sb}(\mu_w)$ holds to a good approximation.

The decay rate for $B\to Ka$ is given by
 \begin{align}\label{eq:b-to-ka}
        \Gamma(B\to K a) &= \frac{m_B}{64\pi}\frac{|C_{sb}(m_b)|^2}{f_a^2}\, f_0^2\left(m_a^2\right)\left(1-\frac{m_K^2}{m_B^2}\right)^2\lambda^{1/2}(m_B^2,m_K^2,m_a^2)\,,
 \end{align}
with the kinematic function $\lambda(a,b,c) =a^2+b^2+c^2 -2(ab + ac + bc)$.  The scalar form factor $f_0\left(m_a^2\right)$ parametrizes hadronic $B\to K$ transitions at momentum transfer $q^2=m_a^2$~\cite{Gubernari:2018wyi}. Expressed in terms of ALP parameters at $\Lambda = 4\pi\,$TeV, the branching ratio for $B^+\to K^+ a$ reads
\begin{align}\label{eq:b-to-ka-num}
    \mathcal{B}(B^+\to K^+ a) = & \ 0.1\left|c_{\ff} - 0.0030\,c_{GG} - 0.0016\,c_{WW}\right|^2\left(\frac{\rm TeV}{f_a}\right)^2\\\nonumber
    & \cdot\frac{f_0^2(m_a^2)}{f_0^2(0)}\frac{\lambda^{1/2}(m_B^2,m_K^2,m_a^2)}{m_B^2-m_K^2}\,.
\end{align}
The decay width of the ALP depends on its mass. For $m_a < 3 m_\pi$ and specifically in the benchmark scenario with $m_a = 300\,$MeV, hadronic ALP decays are kinematically forbidden. The total decay width of the ALP is then given by
\begin{align}\label{eq:lifetime}
\Gamma_a = \sum_{\ell = e,\mu}\Gamma(a\to \ell^+\ell^-) + \Gamma(a\to \gamma\gamma)\,.
\end{align}
The partial widths for decays into leptons and photons are~\cite{Bauer:2017ris,Bauer:2020jbp}
\begin{align}\label{eq:ALP-decays}
        \Gamma(a \to \ell^+\ell^-) & = \left|c_{\ell\ell}(m_a)\right|^2 \frac{m_a m_\ell^2}{8\pi f_a^2}\,\sqrt{1-\frac{4m_\ell^2}{m_a^2}}\,,\\\nonumber
        \Gamma(a \to \gamma\gamma) & = \left|C_{\gamma\gamma}^{\rm eff}(m_a)\right|^2 \left(\frac{\alpha}{4\pi}\right)^2 \frac{m_a^3}{4 f_a^2}\,.
\end{align}
The branching ratio $\mathcal{B}(a \to \mu^+\mu^-)$ depends on the mass $m_a$ and on the relative size of the couplings $c_{ee}$, $c_{\mu\mu}$ and $C_{\gamma\gamma}^{\rm eff}$. For flavor-universal couplings $c_{\ell\ell} = c_{\ff}$, ALP decays into electrons are suppressed compared to muons by the small electron mass. The effective ALP-photon coupling has been calculated in chiral perturbation theory, yielding~\cite{Bauer:2020jbp,Bauer:2021mvw}
\begin{align}\label{eq:cgaga}
    C_{\gamma\gamma}^{\rm eff}(m_a) \approx & \ c_{\gamma\gamma} + c_{ee}(m_a) + c_{\mu\mu}(m_a) - \left(\frac{5}{3} + \frac{m_\pi^2}{m_\pi^2 - m_a^2}\frac{m_d - m_u}{m_d + m_u}\right)c_{GG}\\\nonumber
    & - \frac{1}{2}\frac{m_a^2}{m_\pi^2 - m_a^2}\big(c_{uu}(\mu_\chi) - c_{dd}(\mu_\chi)\big)\,,
\end{align}
This expression applies for ALP masses below the cutoff scale of chiral perturbation theo\-ry, $\mu_\chi \approx 1\,$GeV, and holds to good approximation for $m_a \gg m_\mu$.\footnote{For our numerical analysis, we have implemented the expressions for $c_{\ell\ell}(m_a)$ and $C_{\gamma\gamma}^{\rm eff}(m_a)$ from Ref.~\cite{Bauer:2020jbp}. Furthermore, we set $\mu_\chi = 1\,$GeV and neglect contributions of the strange quark, which are suppressed as $m_{u,d}/m_s$ compared to the up- and down-quark contributions $c_{uu}$ and $c_{dd}$.} For flavor-universal couplings, the total decay width of the ALP depends on the three parameters $c_{\ff}$, $c_{GG}$ and $c_{\gamma\gamma} = c_{WW} + c_{BB}$.
\subsection{Prompt, displaced, invisible}
\label{sec:flavor-signals}
The experimental signature of the process $B\to K a,\,a\to\mu^+\mu^-$ depends on the ALP's lifetime $\tau_a = 1/\Gamma_a$ and on the position and geometry of the detector. The probability to find an ALP, produced at $r = 0$ with boost factor $\beta \gamma$, at a distance $r$ from the production point is given by
\begin{align}
P(r;\beta\gamma) = \exp\left(- \frac{r}{\beta\gamma c\tau_a}\right).
\end{align}
The possible signatures of ALP decays can be classified into three categories: prompt, displaced and invisible. For a prompt signature, the ALP decays within a  distance $r < r_{\rm min}$, where $r_{\rm min}$ is the minimum distance that can be experimentally resolved. The number of observed prompt muon pairs in an experiment is given by
\begin{align}
N_{\rm prompt}(r_{\rm min}) = \mathcal{B}(a\to \mu^+\mu^-) \int d \Phi\, N_a(\vec{r}; \beta\gamma)\,\big[1 - P(r_{\rm min}; \beta\gamma)\big],
\end{align}
where $N_a(\vec{r}; \beta\gamma)$ is the number of ALPs produced with boost $\beta \gamma$ in a small volume around the coordinate point $\vec{r}$; and $d\Phi$ is a phase-space integral over the detectable region.

Displaced ALPs are observed if the decay occurs between $r_{\rm min}$ and the outer detector boundary $r_{\rm max}$. The number of displaced di-muon pairs is
\begin{align}
N_{\rm disp.}(r_{\rm min},r_{\rm max}) = \mathcal{B}(a\to \mu^+\mu^-) \int d \Phi\, N_a(\vec{r}; \beta\gamma)\,\big[P(r_{\rm min}; \beta\gamma) - P(r_{\rm max}; \beta\gamma)\big].
\end{align}
Finally, if the ALP decays outside the detector, the signature is missing energy. The number of events with invisible ALPs is given by\footnote{ALP decays can also appear invisible due to a limited angular detector acceptance and/or detection efficiency.}
\begin{align}\label{eq:ninv}
N_{\rm inv.}(r_{\rm max}) = \int d \Phi\, N_a(\vec{r}; \beta\gamma)\,P(r_{\rm max}; \beta\gamma).
\end{align}
Prompt, displaced and invisible ALP signatures all depend on different combinations of the effective couplings $\{c_{\ff},c_{GG},c_{WW},c_{BB}\}$. The production rate $N_a$ is determined by $\mathcal{B}(B\to Ka) \propto |c_{\ff} - 0.0030\,c_{GG} - 0.0016\,c_{WW}|^2$; the branching ratio $\mathcal{B}(a\to \mu^+\mu^-)$ and decay probability $P(r;\beta\gamma)$ depend on different combinations of $\{c_{\ff},c_{GG},c_{WW} + c_{BB}\}$ through the partial decay rates and the lifetime of the ALP. Due to the exponential decay probability of the ALP, in principle all three signatures coexist for fixed mass and coupling parameters. The relative event rates for these signatures depend on the lifetime of the ALP. Combining searches for prompt, displaced and invisible ALP signals means scanning through the parameter space of effective ALP couplings. We will see that exploiting the lifetime dependence of the ALP signatures greatly helps in resolving the full parameter space of the effective theory.

We demonstrate the complementarity of $B\to K$ signatures numerically by combining searches for prompt and displaced di-muon pairs from $B \to K(\mu^+\mu^-)$ decays by LHCb and a search for $B\to K \me$ with missing energy $\me$ by BaBar. Here we briefly describe the experimental searches; the interpretation of the results in terms of ALP couplings will be presented in Sec.~\ref{sec:flavor-impact}.

\paragraph{Prompt} Short-lived ALPs decay close to their production point. If the decay vertex cannot be resolved by the detector, the events fall within the signal region of $B \to K \mu^+\mu^-$ decays through weak interactions. The LHCb collaboration has performed a search for new scalars $\chi$ in $B^0\to K^{0\ast}\chi, \chi\to \mu^+\mu^-$ decays, allowing the scalar to decay displaced, but not requesting it in the analysis~\cite{LHCb:2015nkv}. Using a total of $3\,$fb$^{-1}$ of data collected at $\sqrt{s} = 7\,$TeV and $8\,$TeV, LHCb has set bounds on the branching ratio for $B^0 \to K^{0\ast} \chi,\,\chi\to \mu^+\mu^-$, which can be directly reinterpreted for ALPs. For $m_a = 300\,$MeV, we obtain the $95\%$ CL upper bound as
\begin{align}\label{eq:lhcb-bound}
\mathcal{B}(B^0 \to K^{0\ast} a)\,\mathcal{B}(a \to \mu^+\mu^-) \lesssim 10^{-9}\qquad \text{for } \tau_a < 10\,\text{ps}.
\end{align}

\paragraph{Displaced} ALPs with intermediate decay lengths can be reconstructed as displaced vertices from their charged decay products. In a similar analysis as described above, LHCb collaboration has performed a dedicated search for displaced di-muon pairs from $B^+\to K^+\chi, \chi\to \mu^+\mu^-$ decays, this time optimizing the search for decay times in three different regions~\cite{LHCb:2016awg}. The bounds on $\mathcal{B}(B^+ \to K^+ \chi)\,\mathcal{B}(\chi \to \mu^+\mu^-)$ are reported for lifetimes in the range $0.1 < \tau < 1259\,$ps. Within this range, the sensitivity is slightly higher than from the analysis in Ref.~\cite{LHCb:2015nkv}. For ALPs with $0.1 < \tau < 1259\,$ps, we therefore also apply the bounds from Ref.~\cite{LHCb:2016awg}.

\paragraph{Invisible} For very long lifetimes, a large fraction of ALPs decays outside the detector and appears as missing energy. The BaBar collaboration has analyzed the events from a search for $B \to K^{(\ast)} \nu\bar{\nu}$~\cite{BaBar:2013npw} in bins of $s_B = q^2/m_B^2$, where $q^2$ is the squared momentum carried by the neutrino pair. A resonant invisible ALP from $B \to K^{(\ast)}a$ would enhance the event rate in the bin containing $s_B = m_a^2/m_B^2$ by $N_{\rm inv.}$, see~\eqref{eq:ninv}. To derive limits on the ALP parameter space, we follow the procedure described in Sec.~3.3 of Ref.~\cite{Ferber:2022rsf} and require that the sum of ALP plus neutrino events falls within 1.64 standard deviations of the background-corrected event rate per bin.

\subsection{Impact of flavor observables on global fit}
\label{sec:flavor-impact}
For the flavor observables, only experimental \emph{bounds} are available, rather than \emph{measurements} as for the high-energy observables. This prevents us from performing a joint fit with high-energy and flavor observables. Instead, we scan the ALP contributions to flavor observables over the region in the $\{c_{\ff},c_{GG},c_{WW},c_{BB}\}$ parameter space that is not excluded by the LHC-LEP fit from Sec.~\ref{sec:comb-lhc}. In this way, we identify regions of parameter points that satisfy the experimental bounds on rare $B$ decays and agree with the measurements of high-energy observables.

\begin{figure}[t!]
    \centering
        \includegraphics[width=0.495\textwidth,page=1]{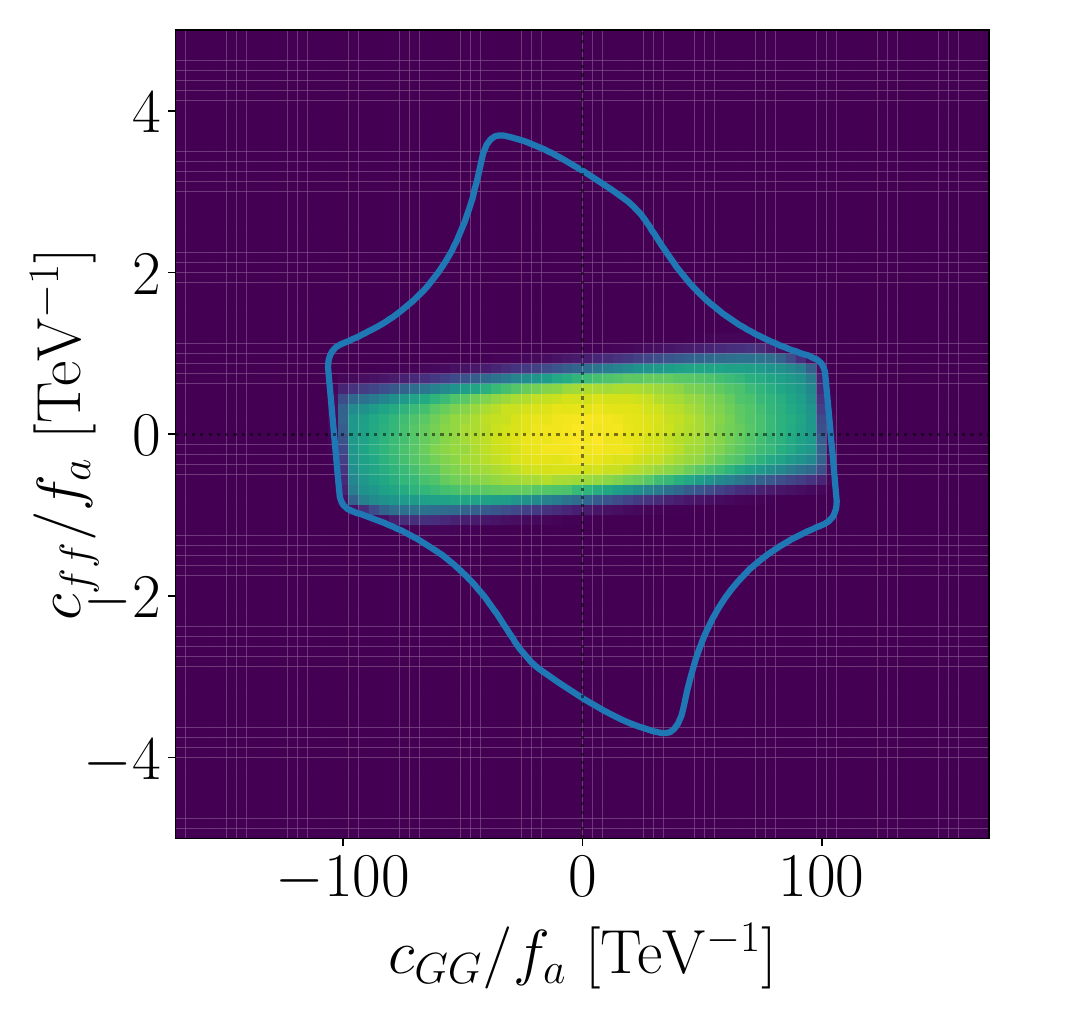}
        \includegraphics[width=0.495\textwidth,page=1]{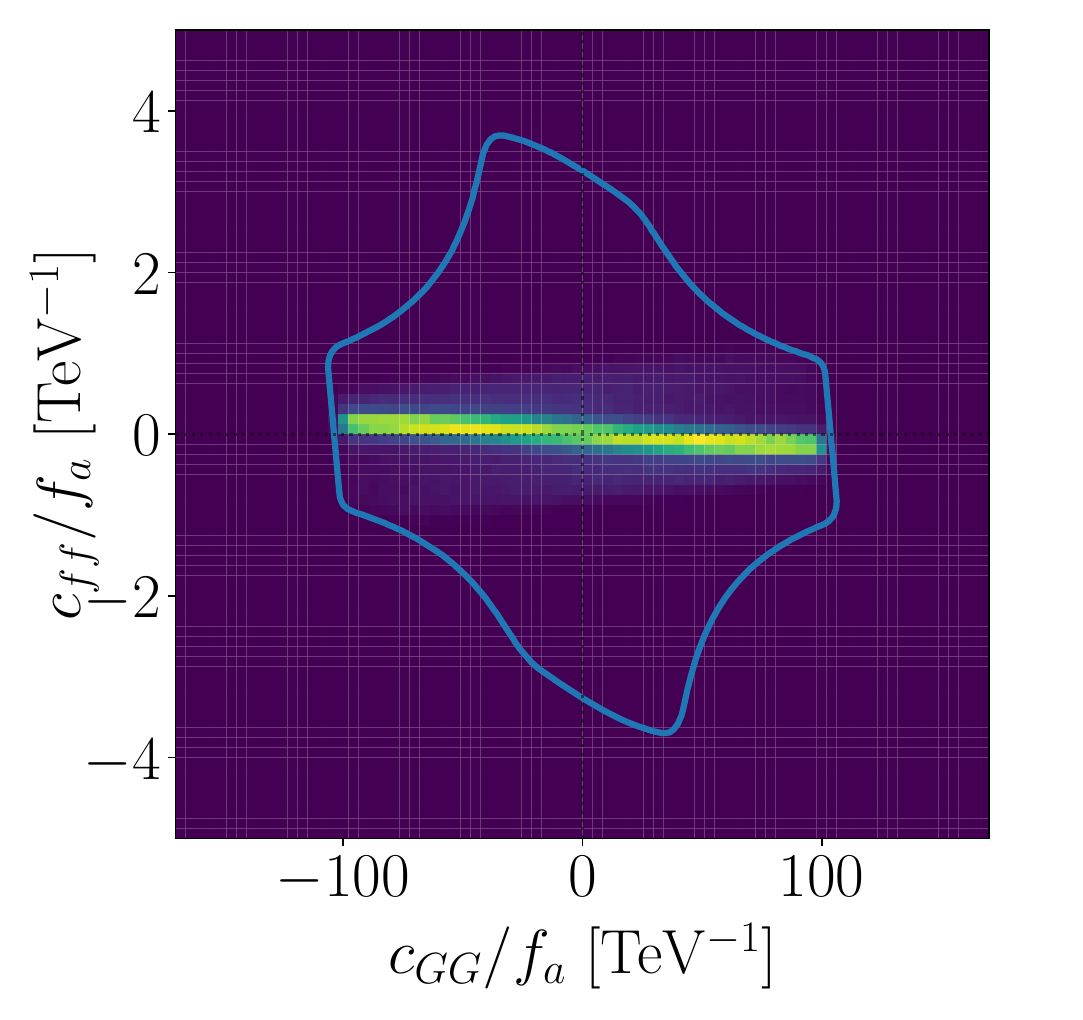}\\
        \includegraphics[width=0.495\textwidth,page=6]{figures/LHC-BaBar.pdf}
        \includegraphics[width=0.495\textwidth,page=6]{figures/LHC-BaBar-LHCb.pdf}
     \caption{Impact of flavor observables in a global analysis of the ALP parameter space $\{c_{\ff},c_{GG},c_{WW},c_{BB}\}$ for $m_a = 300\,$MeV. Shown are projections onto 2-dimensional planes for selected combinations of ALP couplings $c_X(\Lambda)$, defined at the cutoff scale $\Lambda = 4\pi\,$TeV. Left: LHC + $B \to K \slashed{E}$ (BaBar)~\cite{BaBar:2013npw}. Right: LHC + $B \to K \slashed{E}$ (BaBar)~\cite{BaBar:2013npw} + $B \to K (\mu^+\mu^-)$ (LHCb)~\cite{LHCb:2015nkv,LHCb:2016awg}. The blue contours correspond to $\Delta \chi^2 = 5.99$ in a (profiled) 4-parameter fit to high-energy observables (as in Fig.~\ref{fig:global}). The green-yellow areas show viable parameter points obtained from a scan including bounds from the flavor observables. The dark regions are disfavored by the combination of high-energy and flavor observables. A small region around (0,0) is always allowed, but not visible in all panels. All parameters $c_X$ are defined at the cutoff scale $\Lambda = 4\pi\,$TeV.\label{fig:LHC-flavor}}
\end{figure}

In Fig.~\ref{fig:LHC-flavor}, we show the results of this procedure as projections of the 4-dimensional parameter space onto two selected combinations of ALP couplings. The remaining combinations can be found in Fig.~\ref{fig:LHC-flavor-full} in the appendix. The two left panels show the effect of $B \to K \slashed{E}$ on the global analysis; in the two right panels we have added $B\to K(\mu^+\mu^-)$. The results are displayed in terms of ALP couplings $c_X(\Lambda)$ at the cutoff scale $\Lambda = 4\pi\,$TeV.
 The yellow-green areas show the parameter points from our scan that satisfy the high-energy measurements and the bounds from the flavor observables. The color gradient represents the statistical density of points following the LHC-LEP likelihood, from which we have subtracted those points that do not satisfy the bounds from flavor observables.\footnote{The LHC-LEP fit contains more than $2\cdot 10^8$ parameter points, enough to cover the whole parameter space even when many points are removed due to bounds on rare $B$ decays.}
  The dark regions are disfavored by the combination of high-energy and flavor observables. The region around (0,0) is always allowed. For some parameter combinations its size is however so small that it is not visible in the figure.
  
From these plots, it is apparent that the $B\to K$ observables play a big role in resolving the ALP parameter space. The shape of the additional constraints from flavor observables is an intricate interplay of prompt, displaced and invisible  $B\to Ka$ decays, which deserves a detailed discussion.

\paragraph{ALP couplings to fermions and gluons} In the $(c_{GG},c_{\ff})$ plane shown in Fig.~\ref{fig:LHC-flavor} (top row), $B\to K$ observables significantly strengthen the bounds on $c_{\ff}$, while leaving $c_{GG}$ basically untouched. At the boundaries of the favored yellow-green region, the lifetime of the ALP is short enough for the majority of ALPs to decay inside the detector boundaries. The bounds from $B\to K \slashed{E}$ (upper left panel) are therefore due to ALP decays near the production point, which appear invisible because they are emitted outside the detector coverage, for instance near the beam line. In this case, the search for $B\to K \slashed{E}$ is still sensitive, though less than for long-lived ALPs~
\cite{Ferber:2022rsf}. For couplings within the yellow-green region, the $B\to K a$ rate is too small for the bound to apply and the search for $B\to K \slashed{E}$ loses sensitivity. The slight correlation between $c_{\ff}$ and $c_{GG}$ is due to RG effects of $c_{GG}$ in the running of the ALP-top coupling that enter the Wilson coefficient $C_{sb}$, see \eqref{eq:csb-coupling}.

When adding bounds on prompt and displaced $B\to K a, a\to \mu^+\mu^-$ decays (upper right panel), the parameter space is constrained even further. Indeed, searches for (displaced) muon pairs at LHCb are sensitive to smaller event rates than the BaBar search for $B\to K \slashed{E}$. The correlation between $c_{\ff}$ and $c_{GG}$ is different than for $B\to K \slashed{E}$, due to the additional impact of $c_{GG}$ on the ALP decay width, which enters the branching ratio into muons and the decay length of the ALP. Interestingly, a few parameter points are not constrained by di-muon searches, but still by $B\to K \slashed{E}$. This is the case if the ALP branching ratio to muons vanishes due to a subtle interplay of tree-level and loop-induced contributions to $c_{\ff}(m_a)$ from various ALP couplings $c_X(\Lambda)$.

\paragraph{ALP couplings to electroweak gauge bosons} Compared to $c_{\ff}$, the overall sensitivity of $B\to K$ observables to the gauge couplings is much lower, mostly due to the sub-leading effect of $c_{GG}$, $c_{WW}$ and $c_{BB}$ on the ALP production and decay rates. In the $(c_{BB},c_{WW})$ plane in Fig.~\ref{fig:LHC-flavor} (bottom row), $B\to K \slashed{E}$ gives almost no additional information on the ALP parameter space (lower left panel). An exception is the faint darker line along $c_{WW} = - c_{BB}$, where the ALP coupling to photons is suppressed. Since $c_{\ff}$ and $c_{GG}$ are constrained by high-energy observables, the ALP has a long lifetime and tends to decay outside the detector. Along this direction, the scan thus finds less parameter points that satisfy the bounds on $B\to K \slashed{E}$ from BaBar.

 Adding prompt and displaced signatures (lower right panel), however, does remove viable parameter space. As for $B\to K \slashed{E}$, the direction along $c_{WW} = - c_{BB}$ is also disfavored by displaced $B\to K a, a\to \mu^+\mu^-$ decays, since most ALPs decay away from the production point. The orthogonal direction along $c_{WW} = c_{BB}$ is less constrained. Here the ALP decays close to the production point, and the (less sensitive) search for prompt muon pairs applies. As mentioned before, the viable SM-like scenario in a small region around (0,0) is not visible in the figure, due to the small number of 4-parameter combinations in our scan that fall into this region.

\paragraph{Resolving the ALP parameter space} These examples demonstrate the power of parameter correlations in prompt, displaced and invisible ALP decays and their impact on the global analysis of effective ALP couplings. Besides the blind direction along $C_{sb} = 0$, where the production rate vanishes, $B\to K a$ signatures alone can resolve the ALP parameter space. In combination with high-energy observables, all directions are resolved and the sensitivity to individual couplings increases.

We have performed this combination for an ALP with mass $m_a = 300\,$MeV. Since most flavor observables rely on ALP resonances and the decay modes of the ALP depend on its mass, the corresponding bounds are also mass-dependent. For ALP masses below about $10\,$GeV, the energy scale of the $B$ factories, flavor observables will always play a significant role in constraining the ALP parameter space. For larger masses, high-energy observables can be competitive or superior in sensitivity, especially for gauge couplings where the ALP contributions are enhanced at high momenta.

In $B\to K a$ production and ALP decays, we have encountered ALP effects from the RG evolution of the effective theory to energies below the cutoff scale. These effects can generally be sizeable in low-energy observables and offer a largely model-independent way to probe the ALP couplings~\cite{Bauer:2017ris,Alonso-Alvarez:2018irt,Bonilla:2021ufe,Biekotter:2023mpd}. However, RG-induced ALP effects often coincide with tree-level or other loop-induced effects in observables, as we have shown in this study. Depending on the observable, either class of effects can dominate the phenomenology. Their interplay should be analyzed with care.

\section{Conclusions and outlook}
\label{sec:conclusions}
We have presented a combined analysis of high-energy and flavor observables that probes the couplings of the ALP effective theory. Using a selected set of LHC observables in dijet, top-quark and di-boson production, as well as the $Z$ boson width measured at LEP, we resolve all directions in the 4-dimensional parameter space of an ALP with flavor-universal couplings to fermions. We find that dijet, top and $Z$-pole observables are sensitive to non-decoupling effects of the axial vector anomaly in QCD, which modify the sensitivity to ALP couplings compared to tree-level predictions. This characteristic feature of the ALP effective theory may be used to explore the flavor structure of the ALP couplings and to potentially distinguish it from generic scalars or pseudo-scalars.

Low-energy observables drastically improve the resolution of the ALP parameter space. For resonantly produced ALPs in $B\to Ka$ decays, we show that the lifetime of the ALP plays a crucial role: Signatures with prompt, displaced and invisible ALP decays probe complementary directions in the parameter space and help to access even very small coup\-lings. Besides resonance and lifetime effects, low-energy observables are also sensitive to loop effects induced through the running of the ALP couplings. The phenomenology of a low-energy observable is generally determined by all three ALP effects; their relative impact depends on the observable.

Within the effective theory, the various ALP effects in observables at different energies are connected through the renormalization group. This feature allows us to extract the effective ALP couplings, defined at a particular scale, from a combined analysis of high-energy and low-energy observables. The interpretation of such an analysis, however, often depends on the ALP mass, because many observables rely on resonances. This is an important difference with global SMEFT fits, which do not involve new light particles and only probe local effective couplings. Resonances and also most virtual effects of ALPs cannot be mapped onto local effective operators if the momentum scale of the observable is higher than the ALP mass. Effective theories with new light particles describe a particular class of new physics, which requires and deserves its own systematic exploration.

\acknowledgments
We thank Ilaria Brivio for helpful discussions about di-boson production at the LHC. The research of SB and SW is supported by the German Research Foundation (DFG) under grant no. 396021762--TRR 257. The authors acknowledge support by the state of Baden-Württemberg through bwHPC and the German Research Foundation (DFG) through grant no INST 39/963-1 FUGG (bwForCluster NEMO).

\clearpage

\appendix
\section{Global bounds on the ALP effective theory}
\begin{figure}[h!]
    \centering
        \includegraphics[width=0.495\textwidth,page=2]{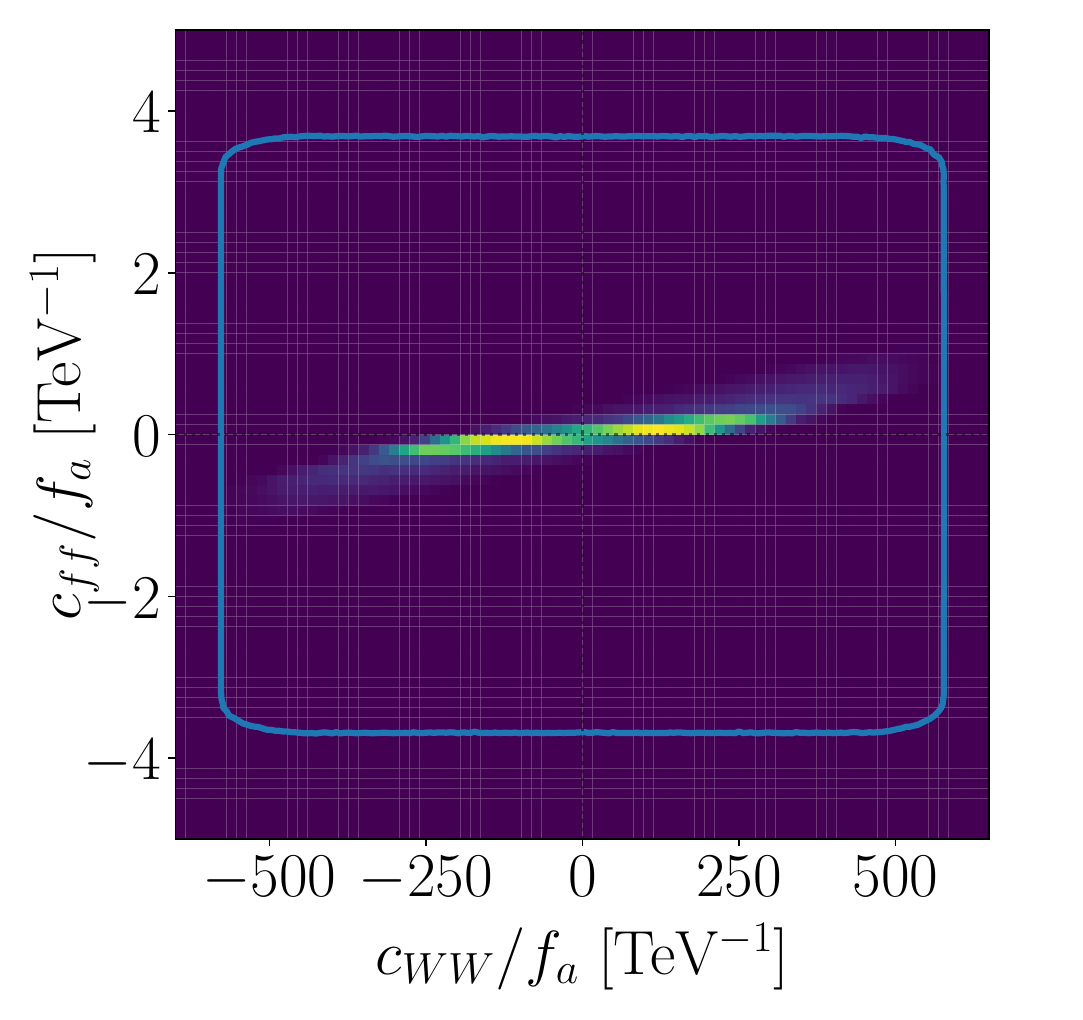}
        \includegraphics[width=0.495\textwidth,page=3]{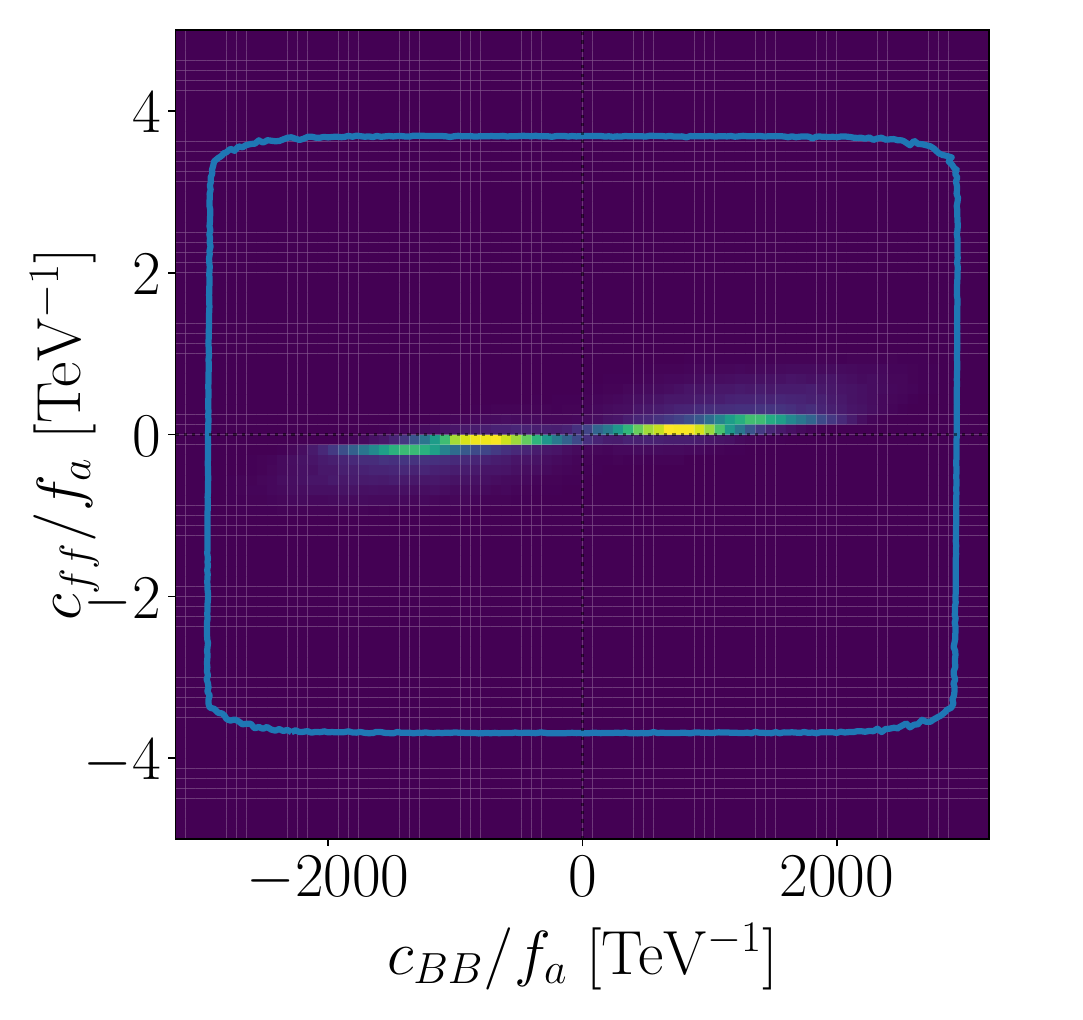} \\
        \includegraphics[width=0.495\textwidth,page=4]{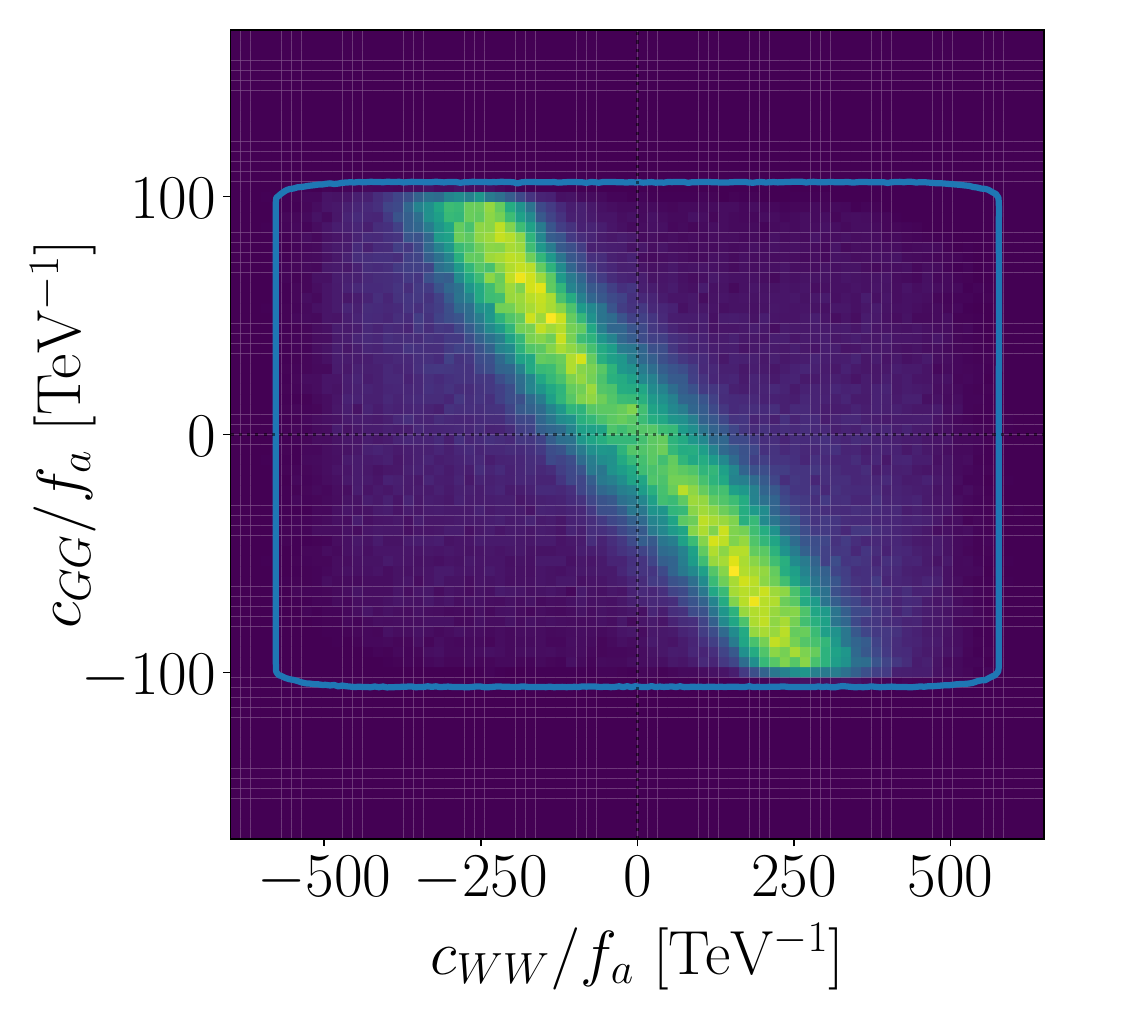}
        \includegraphics[width=0.495\textwidth,page=5]{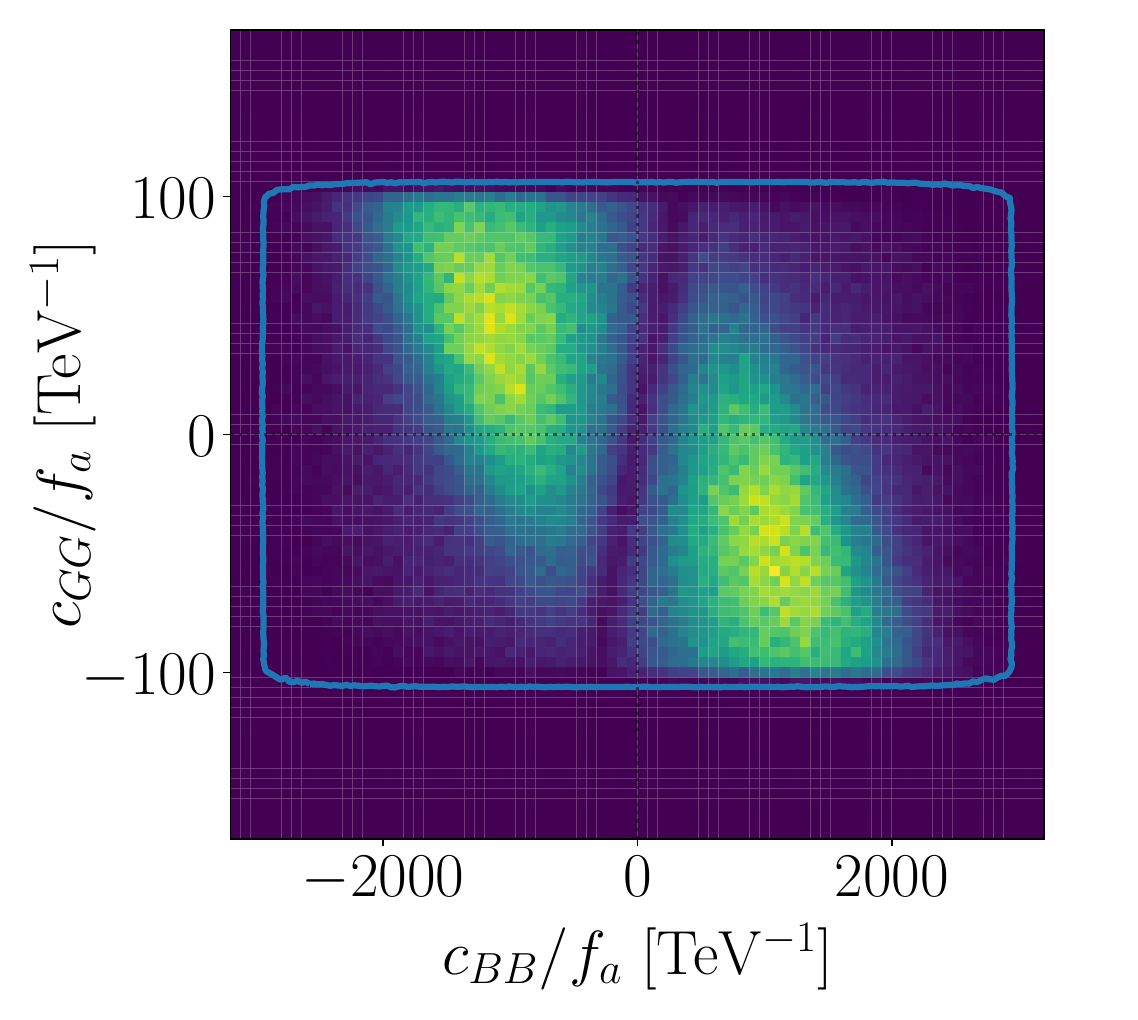}
     \caption{Bounds on the ALP parameter space for $m_a = 300\,$MeV from high-energy observables (blue contours) and parameter regions favored by high-energy plus flavor observables (yellow-green areas). The results are obtained from the same analysis as those in Fig.~\ref{fig:LHC-flavor}, but are displayed for different combinations of ALP couplings.\label{fig:LHC-flavor-full}}
\end{figure}

In this appendix, we provide bounds on the ALP effective theory in 2-dimensional projections of the 4-parameter space. The results in Fig.~\ref{fig:LHC-flavor-full} complement those in Fig.~\ref{fig:LHC-flavor}; both are obtained in the same way described in Sec.~\ref{sec:flavor-impact}.

\clearpage

\bibliographystyle{JHEP_improved}
\bibliography{main}

\providecommand{\href}[2]{#2}\begingroup\raggedright\begin{thebibliography}{10}

\bibitem{Georgi:1986df}
H.~Georgi, D.~B. Kaplan, and L.~Randall,
  \href{http://dx.doi.org/10.1016/0370-2693(86)90688-X}{{\it {Manifesting the
  Invisible Axion at Low-energies}}, } {\em Phys. Lett. B} {\bf 169} (1986)
  73--78.

\bibitem{Peccei:1977hh}
R.~D. Peccei and H.~R. Quinn,
  \href{http://dx.doi.org/10.1103/PhysRevLett.38.1440}{{\it {CP Conservation in
  the Presence of Instantons}}, } {\em Phys. Rev. Lett.} {\bf 38} (1977)
  1440--1443.

\bibitem{Peccei:1977ur}
R.~D. Peccei and H.~R. Quinn,
  \href{http://dx.doi.org/10.1103/PhysRevD.16.1791}{{\it {Constraints Imposed
  by CP Conservation in the Presence of Instantons}}, } {\em Phys. Rev. D} {\bf
  16} (1977) 1791--1797.

\bibitem{Weinberg:1977ma}
S.~Weinberg, \href{http://dx.doi.org/10.1103/PhysRevLett.40.223}{{\it {A New
  Light Boson?}}, } {\em Phys. Rev. Lett.} {\bf 40} (1978) 223--226.

\bibitem{Wilczek:1977pj}
F.~Wilczek, \href{http://dx.doi.org/10.1103/PhysRevLett.40.279}{{\it {Problem
  of Strong $P$ and $T$ Invariance in the Presence of Instantons}}, } {\em
  Phys. Rev. Lett.} {\bf 40} (1978) 279--282.

\bibitem{Alonso-Alvarez:2018irt}
G.~Alonso-\'Alvarez, M.~B. Gavela, and P.~Quilez,
  \href{http://dx.doi.org/10.1140/epjc/s10052-019-6732-5}{{\it {Axion couplings
  to electroweak gauge bosons}}, } {\em Eur. Phys. J. C} {\bf 79} (2019), no.~3
  223, [\href{http://arxiv.org/abs/1811.05466}{{\tt 1811.05466}}].

\bibitem{Baldenegro:2018hng}
C.~Baldenegro, S.~Fichet, G.~von Gersdorff, and C.~Royon,
  \href{http://dx.doi.org/10.1007/JHEP06(2018)131}{{\it {Searching for
  axion-like particles with proton tagging at the LHC}}, } {\em JHEP} {\bf 06}
  (2018) 131, [\href{http://arxiv.org/abs/1803.10835}{{\tt 1803.10835}}].

\bibitem{Gavela:2019cmq}
M.~B. Gavela, J.~M. No, V.~Sanz, and J.~F. de~Troc\'oniz,
  \href{http://dx.doi.org/10.1103/PhysRevLett.124.051802}{{\it {Nonresonant
  Searches for Axionlike Particles at the LHC}}, } {\em Phys. Rev. Lett.} {\bf
  124} (2020), no.~5 051802, [\href{http://arxiv.org/abs/1905.12953}{{\tt
  1905.12953}}].

\bibitem{Carra:2021ycg}
S.~Carra, V.~Goumarre, R.~Gupta, S.~Heim, B.~Heinemann, et~al.,
  \href{http://dx.doi.org/10.1103/PhysRevD.104.092005}{{\it {Constraining
  off-shell production of axionlike particles with Z\ensuremath{\gamma} and WW
  differential cross-section measurements}}, } {\em Phys. Rev. D} {\bf 104}
  (2021), no.~9 092005, [\href{http://arxiv.org/abs/2106.10085}{{\tt
  2106.10085}}].

\bibitem{Bonilla:2022pxu}
J.~Bonilla, I.~Brivio, J.~Machado-Rodr\'\i{}guez, and J.~F. de~Troc\'oniz, {\it
  {Nonresonant Searches for Axion-Like Particles in Vector Boson Scattering
  Processes at the LHC}},  \href{http://arxiv.org/abs/2202.03450}{{\tt
  2202.03450}}.

\bibitem{Mimasu:2014nea}
K.~Mimasu and V.~Sanz, \href{http://dx.doi.org/10.1007/JHEP06(2015)173}{{\it
  {ALPs at Colliders}}, } {\em JHEP} {\bf 06} (2015) 173,
  [\href{http://arxiv.org/abs/1409.4792}{{\tt 1409.4792}}].

\bibitem{Jaeckel:2015jla}
J.~Jaeckel and M.~Spannowsky,
  \href{http://dx.doi.org/10.1016/j.physletb.2015.12.037}{{\it {Probing MeV to
  90 GeV axion-like particles with LEP and LHC}}, } {\em Phys. Lett. B} {\bf
  753} (2016) 482--487, [\href{http://arxiv.org/abs/1509.00476}{{\tt
  1509.00476}}].

\bibitem{Bauer:2017ris}
M.~Bauer, M.~Neubert, and A.~Thamm,
  \href{http://dx.doi.org/10.1007/JHEP12(2017)044}{{\it {Collider Probes of
  Axion-Like Particles}}, } {\em JHEP} {\bf 12} (2017) 044,
  [\href{http://arxiv.org/abs/1708.00443}{{\tt 1708.00443}}].

\bibitem{Brivio:2017ije}
I.~Brivio, M.~B. Gavela, L.~Merlo, K.~Mimasu, J.~M. No, et~al.,
  \href{http://dx.doi.org/10.1140/epjc/s10052-017-5111-3}{{\it {ALPs Effective
  Field Theory and Collider Signatures}}, } {\em Eur. Phys. J. C} {\bf 77}
  (2017), no.~8 572, [\href{http://arxiv.org/abs/1701.05379}{{\tt
  1701.05379}}].

\bibitem{Ebadi:2019gij}
J.~Ebadi, S.~Khatibi, and M.~Mohammadi~Najafabadi,
  \href{http://dx.doi.org/10.1103/PhysRevD.100.015016}{{\it {New probes for
  axionlike particles at hadron colliders}}, } {\em Phys. Rev. D} {\bf 100}
  (2019), no.~1 015016, [\href{http://arxiv.org/abs/1901.03061}{{\tt
  1901.03061}}].

\bibitem{Esser:2023fdo}
F.~Esser, M.~Madigan, V.~Sanza, and M.~Ubiali, {\it {On the coupling of
  axion-like particles to the top quark}},
  \href{http://arxiv.org/abs/2303.17634}{{\tt 2303.17634}}.

\bibitem{Rygaard:2023vyo}
L.~Rygaard, J.~Niedziela, R.~Sch\"afer, S.~Bruggisser, J.~Alimena, et~al., {\it
  {Top Secrets: Long-Lived ALPs in Top Production}},
  \href{http://arxiv.org/abs/2306.08686}{{\tt 2306.08686}}.

\bibitem{Batell:2009jf}
B.~Batell, M.~Pospelov, and A.~Ritz,
  \href{http://dx.doi.org/10.1103/PhysRevD.83.054005}{{\it {Multi-lepton
  Signatures of a Hidden Sector in Rare B Decays}}, } {\em Phys. Rev. D} {\bf
  83} (2011) 054005, [\href{http://arxiv.org/abs/0911.4938}{{\tt 0911.4938}}].

\bibitem{Freytsis:2009ct}
M.~Freytsis, Z.~Ligeti, and J.~Thaler,
  \href{http://dx.doi.org/10.1103/PhysRevD.81.034001}{{\it {Constraining the
  Axion Portal with $B \to K l^+ l^-$}}, } {\em Phys. Rev. D} {\bf 81} (2010)
  034001, [\href{http://arxiv.org/abs/0911.5355}{{\tt 0911.5355}}].

\bibitem{Izaguirre:2016dfi}
E.~Izaguirre, T.~Lin, and B.~Shuve,
  \href{http://dx.doi.org/10.1103/PhysRevLett.118.111802}{{\it {Searching for
  Axionlike Particles in Flavor-Changing Neutral Current Processes}}, } {\em
  Phys. Rev. Lett.} {\bf 118} (2017), no.~11 111802,
  [\href{http://arxiv.org/abs/1611.09355}{{\tt 1611.09355}}].

\bibitem{Choi:2017gpf}
K.~Choi, S.~H. Im, C.~B. Park, and S.~Yun,
  \href{http://dx.doi.org/10.1007/JHEP11(2017)070}{{\it {Minimal Flavor
  Violation with Axion-like Particles}}, } {\em JHEP} {\bf 11} (2017) 070,
  [\href{http://arxiv.org/abs/1708.00021}{{\tt 1708.00021}}].

\bibitem{Bjorkeroth:2018dzu}
F.~Bj\"orkeroth, E.~J. Chun, and S.~F. King,
  \href{http://dx.doi.org/10.1007/JHEP08(2018)117}{{\it {Flavourful Axion
  Phenomenology}}, } {\em JHEP} {\bf 08} (2018) 117,
  [\href{http://arxiv.org/abs/1806.00660}{{\tt 1806.00660}}].

\bibitem{Dobrich:2018jyi}
B.~D\"obrich, F.~Ertas, F.~Kahlhoefer, and T.~Spadaro,
  \href{http://dx.doi.org/10.1016/j.physletb.2019.01.064}{{\it
  {Model-independent bounds on light pseudoscalars from rare B-meson decays}},
  } {\em Phys. Lett. B} {\bf 790} (2019) 537--544,
  [\href{http://arxiv.org/abs/1810.11336}{{\tt 1810.11336}}].

\bibitem{Gavela:2019wzg}
M.~B. Gavela, R.~Houtz, P.~Quilez, R.~Del~Rey, and O.~Sumensari,
  \href{http://dx.doi.org/10.1140/epjc/s10052-019-6889-y}{{\it {Flavor
  constraints on electroweak ALP couplings}}, } {\em Eur. Phys. J. C} {\bf 79}
  (2019), no.~5 369, [\href{http://arxiv.org/abs/1901.02031}{{\tt
  1901.02031}}].

\bibitem{MartinCamalich:2020dfe}
J.~Martin~Camalich, M.~Pospelov, P.~N.~H. Vuong, R.~Ziegler, and J.~Zupan,
  \href{http://dx.doi.org/10.1103/PhysRevD.102.015023}{{\it {Quark Flavor
  Phenomenology of the QCD Axion}}, } {\em Phys. Rev. D} {\bf 102} (2020),
  no.~1 015023, [\href{http://arxiv.org/abs/2002.04623}{{\tt 2002.04623}}].

\bibitem{Carmona:2021seb}
A.~Carmona, C.~Scherb, and P.~Schwaller,
  \href{http://dx.doi.org/10.1007/JHEP08(2021)121}{{\it {Charming ALPs}}, }
  {\em JHEP} {\bf 08} (2021) 121, [\href{http://arxiv.org/abs/2101.07803}{{\tt
  2101.07803}}].

\bibitem{Bauer:2021mvw}
M.~Bauer, M.~Neubert, S.~Renner, M.~Schnubel, and A.~Thamm, {\it {Flavor probes
  of axion-like particles}},  \href{http://arxiv.org/abs/2110.10698}{{\tt
  2110.10698}}.

\bibitem{Ferber:2022rsf}
T.~Ferber, A.~Filimonova, R.~Sch\"afer, and S.~Westhoff, {\it {Displaced or
  invisible? ALPs from $B$ decays at Belle II}},
  \href{http://arxiv.org/abs/2201.06580}{{\tt 2201.06580}}.

\bibitem{Bauer:2019gfk}
M.~Bauer, M.~Neubert, S.~Renner, M.~Schnubel, and A.~Thamm,
  \href{http://dx.doi.org/10.1103/PhysRevLett.124.211803}{{\it {Axionlike
  Particles, Lepton-Flavor Violation, and a New Explanation of $a_\mu$ and
  $a_e$}}, } {\em Phys. Rev. Lett.} {\bf 124} (2020), no.~21 211803,
  [\href{http://arxiv.org/abs/1908.00008}{{\tt 1908.00008}}].

\bibitem{Calibbi:2020jvd}
L.~Calibbi, D.~Redigolo, R.~Ziegler, and J.~Zupan,
  \href{http://dx.doi.org/10.1007/JHEP09(2021)173}{{\it {Looking forward to
  lepton-flavor-violating ALPs}}, } {\em JHEP} {\bf 09} (2021) 173,
  [\href{http://arxiv.org/abs/2006.04795}{{\tt 2006.04795}}].

\bibitem{Dolan:2017osp}
M.~J. Dolan, T.~Ferber, C.~Hearty, F.~Kahlhoefer, and K.~Schmidt-Hoberg,
  \href{http://dx.doi.org/10.1007/JHEP12(2017)094}{{\it {Revised constraints
  and Belle II sensitivity for visible and invisible axion-like particles}}, }
  {\em JHEP} {\bf 12} (2017) 094, [\href{http://arxiv.org/abs/1709.00009}{{\tt
  1709.00009}}]. [Erratum: JHEP 03, 190 (2021)].

\bibitem{Acanfora:2023gzr}
F.~Acanfora, R.~Franceschini, A.~Mastroddi, and D.~Redigolo, {\it {Fusing
  photons into nothing, a new search for invisible ALPs and Dark Matter at
  Belle II}},  \href{http://arxiv.org/abs/2307.06369}{{\tt 2307.06369}}.

\bibitem{Dobrich:2015jyk}
B.~D\"obrich, J.~Jaeckel, F.~Kahlhoefer, A.~Ringwald, and K.~Schmidt-Hoberg,
  \href{http://dx.doi.org/10.1007/JHEP02(2016)018}{{\it {ALPtraum: ALP
  production in proton beam dump experiments}}, } {\em JHEP} {\bf 02} (2016)
  018, [\href{http://arxiv.org/abs/1512.03069}{{\tt 1512.03069}}].

\bibitem{Dobrich:2019dxc}
B.~D\"obrich, J.~Jaeckel, and T.~Spadaro,
  \href{http://dx.doi.org/10.1007/JHEP05(2019)213}{{\it {Light in the beam dump
  - ALP production from decay photons in proton beam-dumps}}, } {\em JHEP} {\bf
  05} (2019) 213, [\href{http://arxiv.org/abs/1904.02091}{{\tt 1904.02091}}].
  [Erratum: JHEP 10, 046 (2020)].

\bibitem{Cadamuro:2011fd}
D.~Cadamuro and J.~Redondo,
  \href{http://dx.doi.org/10.1088/1475-7516/2012/02/032}{{\it {Cosmological
  bounds on pseudo Nambu-Goldstone bosons}}, } {\em JCAP} {\bf 02} (2012) 032,
  [\href{http://arxiv.org/abs/1110.2895}{{\tt 1110.2895}}].

\bibitem{Millea:2015qra}
M.~Millea, L.~Knox, and B.~Fields,
  \href{http://dx.doi.org/10.1103/PhysRevD.92.023010}{{\it {New Bounds for
  Axions and Axion-Like Particles with keV-GeV Masses}}, } {\em Phys. Rev. D}
  {\bf 92} (2015), no.~2 023010, [\href{http://arxiv.org/abs/1501.04097}{{\tt
  1501.04097}}].

\bibitem{Depta:2020wmr}
P.~F. Depta, M.~Hufnagel, and K.~Schmidt-Hoberg,
  \href{http://dx.doi.org/10.1088/1475-7516/2020/05/009}{{\it {Robust
  cosmological constraints on axion-like particles}}, } {\em JCAP} {\bf 05}
  (2020) 009, [\href{http://arxiv.org/abs/2002.08370}{{\tt 2002.08370}}].

\bibitem{Agrawal:2021dbo}
P.~Agrawal et~al.,
  \href{http://dx.doi.org/10.1140/epjc/s10052-021-09703-7}{{\it
  {Feebly-interacting particles: FIPs 2020 workshop report}}, } {\em Eur. Phys.
  J. C} {\bf 81} (2021), no.~11 1015,
  [\href{http://arxiv.org/abs/2102.12143}{{\tt 2102.12143}}].

\bibitem{Chala:2020wvs}
M.~Chala, G.~Guedes, M.~Ramos, and J.~Santiago,
  \href{http://dx.doi.org/10.1140/epjc/s10052-021-08968-2}{{\it {Running in the
  ALPs}}, } {\em Eur. Phys. J. C} {\bf 81} (2021), no.~2 181,
  [\href{http://arxiv.org/abs/2012.09017}{{\tt 2012.09017}}].

\bibitem{Bauer:2020jbp}
M.~Bauer, M.~Neubert, S.~Renner, M.~Schnubel, and A.~Thamm,
  \href{http://dx.doi.org/10.1007/JHEP04(2021)063}{{\it {The Low-Energy
  Effective Theory of Axions and ALPs}}, } {\em JHEP} {\bf 04} (2021) 063,
  [\href{http://arxiv.org/abs/2012.12272}{{\tt 2012.12272}}].

\bibitem{Biekotter:2023mpd}
A.~Biek\"otter, J.~Fuentes-Mart\'\i{}n, A.~M. Galda, and M.~Neubert, {\it {A
  global analysis of axion-like particle interactions using SMEFT fits}},
  \href{http://arxiv.org/abs/2307.10372}{{\tt 2307.10372}}.

\bibitem{Quevillon:2019zrd}
J.~Quevillon and C.~Smith,
  \href{http://dx.doi.org/10.1140/epjc/s10052-019-7304-4}{{\it {Axions are
  blind to anomalies}}, } {\em Eur. Phys. J. C} {\bf 79} (2019), no.~10 822,
  [\href{http://arxiv.org/abs/1903.12559}{{\tt 1903.12559}}].

\bibitem{Bonnefoy:2020gyh}
Q.~Bonnefoy, L.~Di~Luzio, C.~Grojean, A.~Paul, and A.~N. Rossia,
  \href{http://dx.doi.org/10.1007/JHEP07(2021)189}{{\it {The anomalous case of
  axion EFTs and massive chiral gauge fields}}, } {\em JHEP} {\bf 07} (2021)
  189, [\href{http://arxiv.org/abs/2011.10025}{{\tt 2011.10025}}].

\bibitem{Bonilla:2021ufe}
J.~Bonilla, I.~Brivio, M.~B. Gavela, and V.~Sanz,
  \href{http://dx.doi.org/10.1007/JHEP11(2021)168}{{\it {One-loop corrections
  to ALP couplings}}, } {\em JHEP} {\bf 11} (2021) 168,
  [\href{http://arxiv.org/abs/2107.11392}{{\tt 2107.11392}}].

\bibitem{Galda:2021hbr}
A.~M. Galda, M.~Neubert, and S.~Renner,
  \href{http://dx.doi.org/10.1007/JHEP06(2021)135}{{\it {ALP \textemdash{}
  SMEFT interference}}, } {\em JHEP} {\bf 06} (2021) 135,
  [\href{http://arxiv.org/abs/2105.01078}{{\tt 2105.01078}}].

\bibitem{Alloul:2013bka}
A.~Alloul, N.~D. Christensen, C.~Degrande, C.~Duhr, and B.~Fuks,
  \href{http://dx.doi.org/10.1016/j.cpc.2014.04.012}{{\it {FeynRules 2.0 - A
  complete toolbox for tree-level phenomenology}}, } {\em Comput. Phys.
  Commun.} {\bf 185} (2014) 2250--2300,
  [\href{http://arxiv.org/abs/1310.1921}{{\tt 1310.1921}}].

\bibitem{Alwall:2014hca}
J.~Alwall, R.~Frederix, S.~Frixione, V.~Hirschi, F.~Maltoni, et~al.,
  \href{http://dx.doi.org/10.1007/JHEP07(2014)079}{{\it {The automated
  computation of tree-level and next-to-leading order differential cross
  sections, and their matching to parton shower simulations}}, } {\em JHEP}
  {\bf 07} (2014) 079, [\href{http://arxiv.org/abs/1405.0301}{{\tt
  1405.0301}}].

\bibitem{NNPDF:2017mvq}
{\bf NNPDF}, R.~D. Ball et~al.,
  \href{http://dx.doi.org/10.1140/epjc/s10052-017-5199-5}{{\it {Parton
  distributions from high-precision collider data}}, } {\em Eur. Phys. J. C}
  {\bf 77} (2017), no.~10 663, [\href{http://arxiv.org/abs/1706.00428}{{\tt
  1706.00428}}].

\bibitem{Lafaye:2004cn}
R.~Lafaye, T.~Plehn, and D.~Zerwas, {\it {SFITTER: SUSY parameter analysis at
  LHC and LC}},  \href{http://arxiv.org/abs/hep-ph/0404282}{{\tt
  hep-ph/0404282}}.

\bibitem{Lafaye:2007vs}
R.~Lafaye, T.~Plehn, M.~Rauch, and D.~Zerwas,
  \href{http://dx.doi.org/10.1140/epjc/s10052-008-0548-z}{{\it {Measuring
  Supersymmetry}}, } {\em Eur. Phys. J. C} {\bf 54} (2008) 617--644,
  [\href{http://arxiv.org/abs/0709.3985}{{\tt 0709.3985}}].

\bibitem{Lafaye:2009vr}
R.~Lafaye, T.~Plehn, M.~Rauch, D.~Zerwas, and M.~Duhrssen,
  \href{http://dx.doi.org/10.1088/1126-6708/2009/08/009}{{\it {Measuring the
  Higgs Sector}}, } {\em JHEP} {\bf 08} (2009) 009,
  [\href{http://arxiv.org/abs/0904.3866}{{\tt 0904.3866}}].

\bibitem{Bardeen:1969md}
W.~A. Bardeen, \href{http://dx.doi.org/10.1103/PhysRev.184.1848}{{\it
  {Anomalous Ward identities in spinor field theories}}, } {\em Phys. Rev.}
  {\bf 184} (1969) 1848--1857.

\bibitem{Spira:1995rr}
M.~Spira, A.~Djouadi, D.~Graudenz, and P.~M. Zerwas,
  \href{http://dx.doi.org/10.1016/0550-3213(95)00379-7}{{\it {Higgs boson
  production at the LHC}}, } {\em Nucl. Phys. B} {\bf 453} (1995) 17--82,
  [\href{http://arxiv.org/abs/hep-ph/9504378}{{\tt hep-ph/9504378}}].

\bibitem{CMS:2018ucw}
{\bf CMS}, A.~M. Sirunyan et~al.,
  \href{http://dx.doi.org/10.1140/epjc/s10052-018-6242-x}{{\it {Search for new
  physics in dijet angular distributions using proton\textendash{}proton
  collisions at $\sqrt{s}=$ 13 TeV and constraints on dark matter and other
  models}}, } {\em Eur. Phys. J. C} {\bf 78} (2018), no.~9 789,
  [\href{http://arxiv.org/abs/1803.08030}{{\tt 1803.08030}}].

\bibitem{CMS:2021vhb}
{\bf CMS}, A.~Tumasyan et~al.,
  \href{http://dx.doi.org/10.1103/PhysRevD.104.092013}{{\it {Measurement of
  differential $t \bar t$ production cross sections in the full kinematic range
  using lepton+jets events from proton-proton collisions at $\sqrt {s}$ =
  13\,\,TeV}}, } {\em Phys. Rev. D} {\bf 104} (2021), no.~9 092013,
  [\href{http://arxiv.org/abs/2108.02803}{{\tt 2108.02803}}].

\bibitem{Frederix:2017wme}
R.~Frederix, D.~Pagani, and M.~Zaro,
  \href{http://dx.doi.org/10.1007/JHEP02(2018)031}{{\it {Large NLO corrections
  in $t\bar{t}W^{\pm}$ and $t\bar{t}t\bar{t}$ hadroproduction from supposedly
  subleading EW contributions}}, } {\em JHEP} {\bf 02} (2018) 031,
  [\href{http://arxiv.org/abs/1711.02116}{{\tt 1711.02116}}].

\bibitem{CMS:2019jsc}
{\bf CMS}, A.~M. Sirunyan et~al.,
  \href{http://dx.doi.org/10.1007/JHEP11(2019)082}{{\it {Search for the
  production of four top quarks in the single-lepton and opposite-sign dilepton
  final states in proton-proton collisions at $ \sqrt{s} $ = 13 TeV}}, } {\em
  JHEP} {\bf 11} (2019) 082, [\href{http://arxiv.org/abs/1906.02805}{{\tt
  1906.02805}}].

\bibitem{CMS:2019rvj}
{\bf CMS}, A.~M. Sirunyan et~al.,
  \href{http://dx.doi.org/10.1140/epjc/s10052-019-7593-7}{{\it {Search for
  production of four top quarks in final states with same-sign or multiple
  leptons in proton-proton collisions at $\sqrt{s}=$ 13 TeV}}, } {\em Eur.
  Phys. J. C} {\bf 80} (2020), no.~2 75,
  [\href{http://arxiv.org/abs/1908.06463}{{\tt 1908.06463}}].

\bibitem{CMS:2023ica}
{\bf CMS}, {\it {Observation of four top quark production in proton-proton
  collisions at $\sqrt{s}=13\,\mathrm{TeV}$}}, .

\bibitem{ATLAS:2020hpj}
{\bf ATLAS}, G.~Aad et~al.,
  \href{http://dx.doi.org/10.1140/epjc/s10052-020-08509-3}{{\it {Evidence for
  $t\bar{t}t\bar{t}$ production in the multilepton final state in
  proton\textendash{}proton collisions at $\sqrt{s}=13$ $\text {TeV}$ with the
  ATLAS detector}}, } {\em Eur. Phys. J. C} {\bf 80} (2020), no.~11 1085,
  [\href{http://arxiv.org/abs/2007.14858}{{\tt 2007.14858}}].

\bibitem{ATLAS:2021kqb}
{\bf ATLAS}, G.~Aad et~al.,
  \href{http://dx.doi.org/10.1007/JHEP11(2021)118}{{\it {Measurement of the t$
  \overline{t} $t$ \overline{t} $ production cross section in $pp$ collisions
  at $ \sqrt{s} $ = 13 TeV with the ATLAS detector}}, } {\em JHEP} {\bf 11}
  (2021) 118, [\href{http://arxiv.org/abs/2106.11683}{{\tt 2106.11683}}].

\bibitem{ATLAS:2023ajo}
{\bf ATLAS}, G.~Aad et~al., {\it {Observation of four-top-quark production in
  the multilepton final state with the ATLAS detector}},
  \href{http://arxiv.org/abs/2303.15061}{{\tt 2303.15061}}.

\bibitem{ALEPH:2005ab}
{\bf ALEPH, DELPHI, L3, OPAL, SLD, LEP Electroweak Working Group, SLD
  Electroweak Group, SLD Heavy Flavour Group}, S.~Schael et~al.,
  \href{http://dx.doi.org/10.1016/j.physrep.2005.12.006}{{\it {Precision
  electroweak measurements on the $Z$ resonance}}, } {\em Phys. Rept.} {\bf
  427} (2006) 257--454, [\href{http://arxiv.org/abs/hep-ex/0509008}{{\tt
  hep-ex/0509008}}].

\bibitem{ParticleDataGroup:2020ssz}
{\bf Particle Data Group}, P.~A. Zyla et~al.,
  \href{http://dx.doi.org/10.1093/ptep/ptaa104}{{\it {Review of Particle
  Physics}}, } {\em PTEP} {\bf 2020} (2020), no.~8 083C01.

\bibitem{CMS:2020gfh}
{\bf CMS}, A.~M. Sirunyan et~al.,
  \href{http://dx.doi.org/10.1016/j.physletb.2020.135710}{{\it {Measurements of
  production cross sections of WZ and same-sign WW boson pairs in association
  with two jets in proton-proton collisions at $\sqrt{s} =$ 13 TeV}}, } {\em
  Phys. Lett. B} {\bf 809} (2020) 135710,
  [\href{http://arxiv.org/abs/2005.01173}{{\tt 2005.01173}}].

\bibitem{Aiko:2023trb}
M.~Aiko and M.~Endo, \href{http://dx.doi.org/10.1007/JHEP05(2023)147}{{\it
  {Electroweak precision test of axion-like particles}}, } {\em JHEP} {\bf 05}
  (2023) 147, [\href{http://arxiv.org/abs/2302.11377}{{\tt 2302.11377}}].

\bibitem{Gubernari:2018wyi}
N.~Gubernari, A.~Kokulu, and D.~van Dyk,
  \href{http://dx.doi.org/10.1007/JHEP01(2019)150}{{\it {$B\to P$ and $B\to V$
  Form Factors from $B$-Meson Light-Cone Sum Rules beyond Leading Twist}}, }
  {\em JHEP} {\bf 01} (2019) 150, [\href{http://arxiv.org/abs/1811.00983}{{\tt
  1811.00983}}].

\bibitem{LHCb:2015nkv}
{\bf LHCb}, R.~Aaij et~al.,
  \href{http://dx.doi.org/10.1103/PhysRevLett.115.161802}{{\it {Search for
  hidden-sector bosons in $B^0 \!\to K^{*0}\mu^+\mu^-$ decays}}, } {\em Phys.
  Rev. Lett.} {\bf 115} (2015), no.~16 161802,
  [\href{http://arxiv.org/abs/1508.04094}{{\tt 1508.04094}}].

\bibitem{LHCb:2016awg}
{\bf LHCb}, R.~Aaij et~al.,
  \href{http://dx.doi.org/10.1103/PhysRevD.95.071101}{{\it {Search for
  long-lived scalar particles in $B^+ \to K^+ \chi (\mu^+\mu^-)$ decays}}, }
  {\em Phys. Rev. D} {\bf 95} (2017), no.~7 071101,
  [\href{http://arxiv.org/abs/1612.07818}{{\tt 1612.07818}}].

\bibitem{BaBar:2013npw}
{\bf BaBar}, J.~P. Lees et~al.,
  \href{http://dx.doi.org/10.1103/PhysRevD.87.112005}{{\it {Search for $B \to
  K^{(*)} \nu \overline \nu$ and invisible quarkonium decays}}, } {\em Phys.
  Rev. D} {\bf 87} (2013), no.~11 112005,
  [\href{http://arxiv.org/abs/1303.7465}{{\tt 1303.7465}}].

\end{thebibliography}\endgroup

\end{document}